\documentclass[final,3p,times]{elsarticle}

\usepackage{amssymb}
\usepackage{amsmath}

\usepackage{graphicx} 
\usepackage{subfigure}
\usepackage{booktabs} 
\usepackage{siunitx}

\usepackage{hyperref}

\usepackage{etoolbox}
\usepackage{soul} 

\newtoggle{finalPaper}

\setstcolor{blue}

\toggletrue{finalPaper} 

\iftoggle{finalPaper} {
	\newcommand{\addtxt}[1]{#1}
	\newcommand{\change}[2]{#2}
	\newcommand{\rmvtxt}[1]{}}
{
    \soulregister\cite7 
    \soulregister\ref7
    
	\newcommand{\addtxt}[1]{\textcolor{blue}{#1}}
    \newcommand{\change}[2]{\texorpdfstring{\st{#1}\textcolor{blue}{#2}}{#2}}
    \newcommand{\rmvtxt}[1]{\texorpdfstring{\st{#1}}{#1}}
}

\newtoggle{removeItalic}
\togglefalse{removeItalic} 
\iftoggle{removeItalic} {
    \renewcommand{\textit}[1]{#1}
}

\begin{document}

\begin{frontmatter}



\title{Neural cyberattacks applied to the vision under realistic visual stimuli} 

\author[label1]{Victoria Magdalena López Madejska}
\author[label1]{Sergio López Bernal\corref{cor1}}\ead{slopez@um.es}
\author[label1]{Gregorio Martínez Pérez}
\author[label2]{and, Alberto Huertas Celdrán}

\cortext[cor1]{Corresponding author}
\affiliation[label1]{organization={Department of Information and Communications Engineering},
            addressline={University of Murcia, Campus de Espinardo},
            postcode={30100},
            state={Murcia},
            country={Spain}}

\affiliation[label2]{organization={Communication Systems Group CSG, Department of Informatics IfI},
            addressline={University of Zürich UZH},
            postcode={CH—8050},
            state={Zürich},
            country={Switzerland}}



\begin{abstract}
Brain-Computer Interfaces (BCIs) are systems traditionally used in medicine and designed to interact with the brain to record or stimulate neurons. Despite their benefits, the literature has demonstrated that invasive BCIs focused on neurostimulation present vulnerabilities allowing attackers to gain control. In this context, neural cyberattacks emerged as threats able to disrupt spontaneous neural activity by performing neural overstimulation or inhibition. Previous work validated these attacks in small-scale simulations with a reduced number of neurons, lacking real-world complexity. Thus, this work tackles this limitation by analyzing the impact of two existing neural attacks, Neuronal Flooding (FLO) and Neuronal Jamming (JAM), on a complex neuronal topology of the primary visual cortex of mice consisting of approximately 230,000 neurons, tested on three realistic visual stimuli: flash effect, movie, and drifting gratings. Each attack was evaluated over three relevant events per stimulus, also testing the impact of attacking 25\% and 50\% of the neurons. The results, based on the number of spikes and shift percentages metrics, showed that the attacks caused the greatest impact on the movie, while \change{dark and fixed events are the most robust}{dark and static events exhibit highest resilience}. Although both attacks can significantly affect neural activity, JAM was generally more damaging, producing longer temporal delays, and had a larger prevalence. Finally, JAM did not require to alter many neurons to significantly affect neural activity, while the impact in FLO increased with the number of neurons attacked. 

\end{abstract} 

\begin{highlights}
\item Evaluation of two neural cyberattacks on a realistic simulation of the visual cortex.
\item Use of a rich topology with 230,000 neurons from six cortical layers.
\item Attacks tested on different events within three visual stimuli used as visual input.
\item Evaluation of both threats using two metrics: number of spikes and shifts percentage. 
\item Both attacks can disrupt neural activity, having effect for hundreds of milliseconds.
\end{highlights}


\end{frontmatter}



\section{Introduction}
\label{sec:introduction}
Brain-Computer Interfaces BCIs are bidirectional systems that directly interact with the brain, allowing both neural data acquisition and neuromodulation, either by simulating or inhibiting neural activity. These technologies are classified as invasive, if it is necessary to implant electrodes directly into the brain, or non-invasive, if they use an electrode mesh over the scalp \cite{RAMADAN201726}. Non-invasive systems have been used thoroughly in the medical field for the diagnosis of neurological diseases, such as epilepsy \cite{epileptic}. However, these systems have gained special popularity in non-clinical scenarios, such as entertainment and video games, since they are accessible to the general public, present a low cost, and are portable. In contrast, invasive BCIs are useful for stimulating specific brain regions, being relevant for the treatment of neurodegenerative diseases, such as Parkinson's disease \cite{Hartmann2019}. Thus, companies such as Neuralink are focused on creating miniature implantable interfaces capable of offering broad neuronal coverage with low risk during surgery \cite{Musk2019}. In this way, Neuralink currently investigates interfaces able to read and stimulate neurons individually to treat patients with neurological disorders, aiming to democratize the access to neurotechnology in the near future.

Despite the great advantages offered by BCIs, they are not exempt from security and privacy issues. First, the reduction in production costs by manufacturers derive in an increase of the vulnerabilities of these technologies. Moreover, reducing the size of the electrodes in invasive BCIs produce a potential risk on patients' safety during the surgical process. As a result, the literature has focused on documenting the vulnerabilities that affect data confidentiality, data integrity, service availability, and users' safety \cite{Martinovic2012, Li:bciApplications:2015, Ienca:hackingBrain:2016}. In this sense, exploiting these vulnerabilities could damage the correct functionality of these systems, gather users' sensitive data, produce brain damage, or allow attackers to take control of these devices \cite{Lopez_Bernal:cyberBCI:2021}. Based on these vulnerabilities, recent literature has proposed a new family of cyberattacks called neural cyberattacks \cite{Lopez_Bernal:cyberattacks_implants:2020}, which are able to stimulate or inhibit neural activity by assuming that an attacker has already taken control over an invasive neurostimulation BCI with single-neuron resolution. In particular, the authors of \cite{LopezBernal:taxonomy_attacks:2023} introduced a taxonomy of eight neural cyberattacks inspired by well-known attacks in the cybersecurity communications field, capable of disrupting spontaneous neural activity with different malicious behaviors. Those neural cyberattacks were simulated over a non-realistic neuronal topology obtained by training a Convolutional Neural Network (CNN) emulating the visual cortex of mice. This approach was considered due to a lack of realistic neuronal topology during the research. Besides, that work defined metrics able to quantify the impact caused by those cyberattacks regarding neural activity. Addressing the lack of realistic networks existing, recent literature proposed the implementation of neural cyberattacks over a realistic neural topology formed by a small group of 450 neurons \cite{Lopez_Madejska:realitic_topology:2024}. The neuronal model used consists in a detailed reconstruction of the primary visual cortex of mice, specifically from layer four, using static values as visual input. In addition, the literature used existing metrics that quantified the impact of two neural cyberattacks and analyzed the possible effect on the vision. 

Despite the advances towards evaluating neural cyberattacks on realistic neuronal representations, the literature still presents important limitations. First, there is a need to use larger and more complex topologies representing a sufficiently large portion of the visual cortex to obtain more evidence regarding the impact of neural cyberattacks over the brain. Secondly, the literature has validated the impact of neural cyberattacks using visual stimuli lacking realism. Therefore, evaluating multiple realistic visual stimuli as input of the neural simulation is an open challenge. Thirdly, the set of metrics currently explored to assess the impact of these threats is limited and, thus, new approaches are required. Finally, there is a lack of detailed research on the impact of these attacks on critical cognitive functions, such as vision. Based on the previous limitations, the present work intends to advance the state of art by providing new orientations about the impacts that entail neural cyberattacks, thus presenting the following contributions:

\begin{itemize}
    \item Implementation of Neuronal Flooding (FLO) and Neuronal Jamming (JAM), two neural cyberattacks previously defined in the literature, on a neuronal simulation using a reconstruction of the primary visual cortex (V1) of mice formed by around 230,000 neurons, distributed among the six layers of V1. In addition, this work evaluated the impact of the cyberattacks in a realistic way by including three visual stimuli into the simulation, aiming to recreate real-world visual conditions. These stimuli are a flash light, a fragment of a movie, and drifting gratings with an orientation of 90 degrees. This work used two main tools, BMTK and NEST, to build and simulate the model. The neural activations obtained from the simulation are analyzed using two impact metrics from the literature: number of spikes and percentage of shifts. The first metric indicates the number of activations (action potentials or spikes) of all the neurons of the topology over a specific interval, whereas the second shows the percentage of all the spikes delayed in time compared to normal behavior. 
    \item The evaluation of the results obtained from the experiments using the metrics previously mentioned. This research demonstrates that the movie stimulus is the most sensitive against neural cyberattacks, while the dark event in the flash stimulus is the most robust. However, attacking an event where the neurons generate a few spikes in their normal behavior causes a significant increase compared to spontaneous signaling. Moreover, both attacks have a notorious impact based on the number of spikes metric. However, the impact in FLO is determined by the stimulus, whereas JAM is able to inhibit any event, being the most damaging attack, affecting the 99\% of spikes. In addition, the shifts percentage shows that the attacks produce a high delay over the whole simulation, although JAM produces longer delays (from 50\% to 95\%) compared to 40\%-90\% for FLO. Furthermore, the experiments for JAM conclude that it is not necessary to attack a great number of neurons to cause a significant impact on neural activity. On the contrary, the results for FLO indicate that, the higher the number of attacked neurons, the greater the impact. Additionally, the effect of the attacks gets propagated in time when evaluating this metric, where FLO extends its impact around 200-300 ms after the attack instant, while JAM depends on the event, ranging between 600 ms and 1,100 ms according to the situation.
    \item Analysis of the impact caused by the cyberattacks based on specific parameters used to configure their behavior. The first parameter defines the instants where the attacks are active, selecting in this work three instants per stimulus corresponding to the most significant events within each stimulus. First, the experiments tested the impact of attacking over a neutral event, which is common in all three stimuli. Besides, this work evaluated bright and dark events for the flash events, while for the movie and drifting gratings stimuli the attacks were tested at the beginning and the middle of the stimuli, corresponding to different degrees of intensity in spontaneous neural activity. The second parameter studied is the number of neurons attacked over the whole topology, evaluating the impact of attacking the 25\% and 50\% of the total of the neurons. Finally, another important parameter is the voltage used for the attacks, which corresponds to the maximum value of the natural range of the neurons for FLO and the reset value for JAM.
    \item Comparison between the literature and this research. Attending to the neuronal topology, the literature employed 450 neurons with no realistic input, whereas this work used around 230,000 neurons with realistic visual stimuli as input. Secondly, the comparison considers the results of both cyberattack experiments where FLO has a similar conclusion: the higher the number of neurons affected, the higher the number of spikes. However, the conclusions obtained for JAM differ from those within the literature since this work demonstrates that it is not necessary to attack a high number of neurons to produce a notorious impact. Finally, the attack propagation is slightly different: the propagation of FLO in the literature is 500 ms, whereas in this manuscript, the effect caused by FLO is visible around 200-300 ms after the attack. In contrast, the literature documents that the alteration induced by JAM is still active after 600ms, while in this work, this attack achieves a maximum propagation of 1,100 ms until converging with spontaneous signaling. 
  
\end{itemize}

This article is organized as follows. Section \ref{sec:relatedwork} documents the state of the art of cybersecurity on BCIs and the literature regarding neural cyberattacks in different neural topologies. Section \ref{sec:proposedSolution} details the design and the implementation of the proposed solution and its fundamental components. Afterward, Section \ref{sec:experimentationscenario} defines different experimental scenarios used to implement and validate the neural cyberattacks, evaluating the impact of each attack on different events within the three visual stimuli considered. Then, Section \ref{sec:results} shows the results obtained on each experiment, analyzed based on two impact metrics. Section \ref{sec:discussion} compares the results of both cyberattacks executed over different stimuli based on different criteria and, then, with those existing in the literature. Finally, Section \ref{sec:conclusions} presents the conclusions and future work.

\section{Related Work}
\label{sec:relatedwork}
This section first presents \addtxt{the neural models studied from the literature and their characteristics. Next, this section descibes} the state of the art oriented to broad aspects of cybersecurity on BCIs, showing potential attacks on confidentiality, integrity, and availability of BCIs, as well as users' safety. After that, this section presents the works related to neural cyberattacks, highlighting the most relevant works and emphasizing important aspects of the solutions proposed, such as targeted brain areas and neuronal topologies used.

\addtxt{\subsection{Reconstruction of realistic neural models.}}
\addtxt{There are numerous models focused on different aspects of the brain. These models are realistic but on a small scale. Moreover, most of them are not sufficiently complex or complete to study the dynamics of neural cyberattacks against neural activity. Therefore, this work reviews models that realistically simulate and replicate neural behavior, presenting sufficient details. Nevertheless, for the sake of simplicity, this section highlights the most relevant models for this research. The selection criteria focus on models providing a complete reconstruction of a specific brain microcircuit, offering a high degree of biological realism, and if they are publicly available for research purposes. }

\addtxt{Arkhipov et al. \cite{Arkhipov2018} released a biophysically detailed model of layer 4 of mice V1, consisting of 45,000 neurons, containing excitatory and inhibitory cells. This model includes 10,000 biophysical cells and 35,000 leaky-integrate-and-fire (LIF) neurons. The last type of neuron is a simplified model of a biological neuron focused on integrating incoming signals, filtering some of the potential over time, and generating output spikes. These cells were distributed on a cylinder of 100 $\,\mu$m in height. The biophysical cells are located at 400 $\,\mu$m radius of the inner core, whereas the LIF cells surrounded the core from 400 $\,\mu$m to 845 $\,\mu$m. Additionally, the model received a thalamocortical visual input derived from the transformation of visual stimuli into spike by lateral geniculate nucleus (LGN) filters. This work was able to simulate with precision the neural behavior of L4 in response to visual stimuli, highlighting the importance of the connections between them. However, this model presents some limitations, such as a lack of selectivity in direction caused by the simplification of the thalamocortical input.}

\addtxt{Next, Billeh et al. \cite{billeh:topology:2020} presented a biological model of mice V1, representing both excitatory and inhibitory neurons. In particular, it is formed by \textasciitilde{230,000} neurons from models differentiated by their morpho-electrical characteristics and the cortex layer in which they are located. Additionally, the neurons are placed in a cortex microcolumn, with a radius of 845 $\,\mu$m. The neuron distribution followed is based on the previous research by Arkhipov et al. \cite{Arkhipov2018}. This work offers two levels of granularity: the first option utilizes biophysically detailed neurons, while the second employs Gated Leaky Integrate-and-Fire (GLIF) neurons following a point-neuron approach. Both models were tested using different visual stimuli: full-field flashes, a natural movie, and a looming disk, all preceded by a gray screen, representing a neutral visual event. Both approaches exhibited functional similarities, supporting the use of simpler models such as GLIF for large-scale functional studies.}

\addtxt{Aussel et al. \cite{Aussel2018} created a computational model of the healthy hippocampus (CA1, CA3, and dentate gyrus) as well as the entorhinal cortex , which presents characteristic rhythms of wakefulness and slow-wave sleep. This network comprises 33,200 neurons, including 31,000 excitatory neurons and 2,200 inhibitory interneurons, distributed across the following components: CA1 (10,000 excitatory neurons; 1,000 interneurons), CA3 (1,000; 100), dentate gyrus (10,000; 100), and entorhinal cortex (10,000; 1,000). The authors utilized point-neurons with realistic dynamics, and conductance-based Hodgkin-Huxley neurons. The results showed that this model can reproduce two types of oscillations observed in humans: sharp-wave ripples during deep sleep and theta-nested gamma oscillations during wakefulness. Later, the same authors extended the previous model of a healthy hippocampus, including four typical pathological modifications of the hippocampus seen in medial temporal lobe epilepsy \cite{Aussel2022}. This extension aimed to investigate how these structural and functional abnormalities, both at the neuronal and network levels, contribute to the generation of seizure-like theta activity under different brain states (wakefulness and slow-wave sleep).}

\addtxt{Finally, Romani et al. \cite{Romani2024} presented a full-scale model of the rat hippocampus, particularly from CA1, comprising 456,000 neurons derived from 43 morphological reconstructions of neurons belonging to 12 distinct morphological types. The authors focused on the reconstruction and validation of the model, followed by the simulation and analysis of activity in this reconstructed topology. Additionally, the input used in this work is, mainly, the Schaffer collaterals from CA3 pyramidal cells, the most prominent afferent input to the CA1 and the most studied pathway in the hippocampus. They concluded that this model is one of the most detailed and comprehensive of this region to date, capable of reproducing realistic dynamics. However, it still presents limitations, as the model has incomplete structures due to a lack of cell classes, data, and standardized validation.}

\subsection{General cybersecurity issues of BCIs}
As BCIs have gained great popularity in recent years, there has been an increase of works focused on their vulnerabilities, addressing different traditional cybersecurity dimensions. Concerning data confidentiality, Martinovic et al. \cite{Martinovic2012} used malicious visual stimuli to obtain sensitive information from users' brain waves in response to the stimuli, obtaining information related to passwords, credit cards, residence, and religious beliefs, among others. In the same way, Frank et al. \cite{Frank2017} and Quiles Pérez et al. \cite{QuilesPérez2021} studied the impact of presenting malicious visual stimuli with different durations, also exploring the subliminal realm, demonstrating that attackers could use brief stimuli to obtain sensitive information from users unaware of the attack, which affects their privacy. Finally, Takabi et al. \cite{takabi:privacyThreatsCounter:2016} identified the most common vulnerabilities in BCI applications, highlighting that most applications offered unrestricted access to brain data. In addition, Bonaci et al. \cite{bonaci:appStores:2015} presented how to anonymize the signals transmitted from the BCI to external devices, proposing a system called BCI Anonymizer.

Attending to data integrity problems, Li et al. \cite{Li:bciApplications:2015} indicated that attackers could take advantage of the vulnerabilities existing in BCIs to gain access to sensitive data, thus allowing them to modify the information or, even, perform replay attacks by impersonating legitimate brain waves with previously recorded signals. Moreover, Martínez Beltrán et al. \cite{MartínezBeltrán2022} demonstrated that introducing noise into brain signals could disrupt the correct functioning of the classifiers used by BCIs systems to identify relevant events in cerebral activity. Thus, this work confirmed that modifying the data to confuse the classifiers greatly impacts service integrity and availability.

Focusing on service availability, both Li et al. \cite{Li:bciApplications:2015} and Ienca et al. \cite{ienca:bciConsumerDevices:2018} identified the possibility of disrupting the data acquisition process using different attack vectors, potentially leading to system downtime or interruptions in data capture. Such disruptions may affect critical applications where an uninterrupted data flow is essential. Besides, Cámara et al. \cite{Camara:IMDsecurity:2015} presented attacks focused on draining the battery of stimulation systems by flooding them with continuous requests or signals. These attacks exploit vulnerabilities in the power management of such systems, leading to rapid depletion of the battery and reducing device availability, affecting as well users' safety.

Regarding users' safety, the literature has demonstrated that attacks targeting BCIs could damage patients' brains. In this context, Ienca et al. \cite{Ienca:hackingBrain:2016} and Pycroft et al. \cite{pycroft:brainjacking:2016} highlighted that overstimulating the brain could damage brain tissue or cause a rebound effect, thus altering medical treatments. Additionally, Marin et al. \cite{Marin:securityNeurostimulators:2018} emphasized that manipulating these devices to perform incorrect neuromodulation could affect speech or movement, causing brain damage and, even, risking users' lives. Finally, Landau et al. \cite{Landau:security_EEG:2020} indicated that changing diagnostic test results using BCIs could lead to incorrect or unnecessary treatments for patients. 

Despite the various studies focused on the general view of BCI cybersecurity, no study offered a standardized vision of this topic. Therefore, López Bernal et al. \cite{Lopez_Bernal:cyberBCI:2021} comprehensively reviewed the cybersecurity status of BCIs. The authors unified the BCI life-cycle, defined at that moment in different ways in the literature, and offered a study of attacks, impacts, and countermeasures of each stage of the proposed cycle. Also, this work did a similar analysis focusing on common architectural designs in BCI deployments. Finally, the authors highlighted the limitations of the literature and future challenges that should be tackled to increase the security of BCI systems. 

In summary, Table \ref{tableCybersecurityPaper} shows all the previously mentioned works, their approach (neural data acquisition or neurostimulation), the impact dimension they cover (confidentiality, integrity, availability, or safety), and a brief description of each threat. 

\begin{table}[!htb]
\centering
\label{tableCybersecurityPaper}
\resizebox{\columnwidth}{!}{
\begin{tabular}{@{}llll@{}}
\toprule
\textbf{Work/Reference}& \textbf{Approach}& \textbf{Impact}& \textbf{Threat} \\
\midrule
Martinovic et al. \cite{Martinovic2012} (2012) & Acquisition& Confidentiality & Malicious visual stimuli\\ \midrule
Bonaci et al. \cite{bonaci:appStores:2015} (2015) & Acquisition& Confidentiality & Unrestricted applications\\ \midrule
Camara et al. \cite{Camara:IMDsecurity:2015} (2015) & Stimulation& Availability, safety & Brain damage\\ \midrule
Li et al. \cite{Li:bciApplications:2015} (2015) & Acquisition& Integrity, availability & Replace legitimate signals, stop the BCI\\ \midrule
Ienca et al. \cite{Ienca:hackingBrain:2016} (2016) & Stimulation& Safety & Brain damage\\ \midrule
Pycroft et al. \cite{pycroft:brainjacking:2016} (2016) & Stimulation& Safety & Brain damage\\ \midrule
Takabi et al. \cite{takabi:privacyThreatsCounter:2016} (2016) & Acquisition& Confidentiality& Unrestricted applications\\ \midrule
Frank et al. \cite{Frank2017} (2017) & Acquisition& Confidentiality & Malicious visual stimuli\\ \midrule
Ienca et al. \cite{ienca:bciConsumerDevices:2018} (2018) & Acquisition& Availability & Stop the BCI\\ \midrule
Marin et al. \cite{Marin:securityNeurostimulators:2018} (2018) & Stimulation& Safety & Brain damage\\ \midrule
Landau et al. \cite{Landau:security_EEG:2020} (2020) & Acquisition& Safety & Disrupt diagnostic tests\\ \midrule
Quiles Pérez et al. \cite{QuilesPérez2021} (2021) & Acquisition& Confidentiality & Malicious visual stimuli\\ \midrule
López Bernal et al. \cite{Lopez_Bernal:cyberBCI:2021} (2021) & Acquisition, stimulation& All & Comprehensive analysis of several threats\\ \midrule
Martínez Beltrán et al. \cite{MartínezBeltrán2022} (2022) & Acquisition& Integrity, availability & Disrupt brain waves, alter classifiers\\ 
\bottomrule
\end{tabular}}
\caption{Summary of works related to cybersecurity on BCIs, presenting the neuromodulation approach followed, the dimension affected, and the description of each threat covered.}
\end{table}

\subsection{Neural cyberattacks to alter neural activity}

López Bernal et al. \cite{Lopez_Bernal:cyberattacks_implants:2020}, as a continuation of their cybersecurity analysis in \cite{Lopez_Bernal:cyberBCI:2021}, identified vulnerabilities in the design of new-generation neurostimulation systems (e.g., Neuralink \cite{Musk2019}), cyberattackers could exploit that to take control of the behavior of individual neurons. The authors also presented a new concept of cyberattacks called neural cyberattacks, which can individually stimulate or inhibit neurons accessible by a vulnerable BCI following different attacking strategies. This research defined two neural cyberattacks: Neuronal Flooding (FLO) and Neuronal Scanning (SCA). FLO focuses on overstimulating a set of neurons in a specific instant, whereas the functioning of SCA consists in stimulating neurons sequentially over time. The authors opted to train a Convolutional Neural Network (CNN) formed by 276 neurons, divided into three layers, 200 neurons for the first layer, then 72 neurons for the second layer, and the last one has four neurons. This kind of model was selected because of the lack of realistic neural topologies during the research. Moreover, this decision was justified by the similarities that this kind of network has with the biological visual cortex according to its structure and function, as indicated by Kuzovkin et al. \cite{Kuzovkin2018}. This paper trained the network to resolve the specific problem of a mouse that has to solve a certain maze, receiving a simplified visual input related to adjacent cells from its actual position. The connections and weights of the CNN model were translated into simulated biological neural network parameters. Additionally, the authors obtained the optimal path of the maze to exit the maze from the starting position in the CNN, resulting in 27 positions. Then, this work performed a simulation of 27 seconds, emulating that the mouse stayed one second per position.

This work evaluated the FLO impact when it was used to randomly attack different neurons, ranging from five to 105 neurons from the first layer. In contrast, SCA was used to attack 200 neurons from the first layer sequentially. Both attacks were validated and implemented in a constrained topology representing a portion of the visual cortex of mice. To measure the impact of the previous cyberattacks, the authors defined different impact metrics: number of spikes, percentage of shifts, and dispersion of spikes (temporal, spikes). The first metric quantifies the number of activations (or action potentials) of the neurons attacked compared to spontaneous neuronal signaling. The second metric studies the delay of a spike in time (forward or backward) compared to spontaneous activity with the aim of analyzing how cyberattacks delay the normal behavior of neurons. The third metric analyzes the distribution of spikes over time by studying the percentage of instants that have spikes and if the cyberattack causes a modification in the distribution of the spikes. Finally, this publication concluded that SCA caused a greater reduction in the number of spikes compared to FLO; on the contrary, FLO induced a higher percentage of shifts than SCA. Finally, according to the dispersion of spikes, both attacks show similar values. In summary, FLO was more effective in generating immediate dispersion, whereas SCA generated sustained and gradual dispersion. 

Afterward, López Bernal at el. \cite{Lopez_Bernal:jamming:2022} presented Neuronal Jamming (JAM) as a new neural cyberattack capable of inhibiting neural activity during a temporal interval. The aim of the work is to explore the impact it can generate on the brain of this kind of cyberattack. The authors used the previously mentioned topology formed by 276 neurons to validate the cyberattack. The experiments of JAM were configured according to the number of neurons attacked (from five to 105 neurons),  the temporal window (from 10 ms to 60 ms), and the number of consecutive attacked positions of the maze (from one to 27). In addition, the publication used different metrics to measure the impact: the number of spikes and temporal dispersion. The results show that JAM in the model decreases the number of spikes and the temporal dispersion equally. After presenting the experiments for this new attack, the authors compared the impact of JAM with FLO. In this case, FLO affects from five and 105 neurons at the instant 50 ms and the position of the maze individually. These experiments show that FLO increases the number of spikes and temporal dispersion. This work concluded that both cyberattacks generate a notorious impact, but FLO generates a more disruptive impact than JAM. Furthermore, the authors compare this model with the artificial approach with the aim of verifying if there is a relationship between the impact of neurons and the mouse's performance in exiting the maze. The results show that JAM increases the number of steps to reach the exit, decreasing the success rate in solving the problem. On the contrary, FLO increases the number of steps, whereas it decreases the success rate. Therefore, these results indicated a relationship, although this paper is developing and has realism constraints.

Finally, López Bernal et al. \cite{LopezBernal:taxonomy_attacks:2023} defined a taxonomy of eight neural cyberattacks, including the previous three: Neuronal Flooding (FLO), Neuronal Jamming (JAM), Neuronal Scanning (SCA), Neuronal Selective Forwarding (FOR), Neuronal Spoofing (SPO), Neuronal Sybil (SYB), Neuronal Sinkhole (SIN), and Neuronal Nonce (NON). These attacks are differentiated based on the following aspects: 1) malicious actions based on stimulation or inhibition, 2) execution in a specific instant or during a temporal window, 3) low or high complexity, and 4) number of neurons involved per instant. For example, SPO replicates the behavior of a set of neurons during a given period, SYN stimulates neurons from superficial layers connected to neurons placed in deeper layers, and NON focuses on performing neural stimulation, inhibition, or a combination of both actions by targeting a random set of neurons in specific instants. As in the previous papers, each attack was validated in the same constrained topology of 276 neurons. In this case, according to the attack, the number of neurons attacked changes as well as the start of the attack instant, based on the characteristics and inner behavior of each specific cyberattack. This paper uses the number of spikes as a metric to evaluate the impact of the cyberattacks. The results highlighted that JAM and NON have notorious impacts in the short term, decreasing the neural activity over 5\% and 12\%, respectively. In contrast, SCA and NON are more effective in the long term, with a reduction of 9\% and 8\%, respectively. The short term was evaluated over the first five seconds of a simulation of 27 seconds, while the long term was over the last five seconds. 

All the previous publications are limited by the use of a non-realistic neuronal topology. However, in recent years, the literature has provided works related to the reconstruction. Therefore, López Madejska et al. \cite{Lopez_Madejska:realitic_topology:2024} implemented FLO and JAM cyberattacks in a reconstruction of the primary visual cortex (V1) of mice, in particular the fourth layer, published by Arkhipov et al. \cite{Arkhipov2018}. The topology comprises approximately 45,000 neurons with two different levels of detail in the neuronal simulation, one simpler based on point-neurons and the other presenting more biophisiological details. To simplify the complexity of the experimentation, López Madejska et al. chose a smaller topology of 450 neurons provided as well by Arkhipov et al., extracted from the complete model. The authors used the Brain Modeling Toolkit \cite{BMTK} (BMTK) tool developed by the Allen Institute to create and manage the model and the simulator NEST \cite{Gewaltig:NEST} to simulate it. Besides, the authors employed realistic static values represented as Lateral Geniculate Nucleus (LGN) spikes (neuronal activity). LGN is a brain component responsible for processing and forwarding the external stimulus to V1, provided by the BMTK as external visual stimuli. 

The work previously mentioned accomplished two FLO experiments: the first evaluates how different voltages affect a set of neurons (5 mV, 10 mV, 20 mV, 30 mV, and threshold voltage). On the contrary, the second experiment analyzes the effect of attacking different neurons simultaneously using a specific voltage (50, 100, 200, 300, 400, and 450 neurons). Both experiments used the instant 10 ms as the instant under attack. For JAM, the authors studied the impact of attacking different numbers of neurons as the second experiment of FLO during a time window (1,000-1,500 ms). The results for FLO show that forcing the threshold voltage, which guarantees that a targeted neuron will generate a spike, is the most damaging alternative. Besides, the number of spikes increases when attacking more neurons. In contrast, the experimentation for JAM demonstrated that the greater the number of neurons attacked, the lower the number of spikes. Additionally, in both attacks, the neural activity tends to return to spontaneous behavior, consistent with neuroscience evidence. In the case of FLO, the neurons take 500 ms to return to spontaneous behavior, whereas in JAM, the neurons take 600 ms. Moreover, the authors compared their results with the literature, highlighting the difference between neuronal models, the number of neurons affected, and the impact generated by the cyberattacks.

In summary, Table \ref{tableNeuralCyberattacks} presents the aforementioned papers, the neural cyberattacks analyzed, the neural model and metrics used, and a brief description of the results obtained. In a nutshell, existing works focused on neural cyberattacks only tested small topologies, with a maximum of 450 neurons. Moreover, the complexity of the realistic neuronal reconstructions employed is limited, only representing one layer of the visual cortex, lacking sufficient realism to understand the implications within the visual process. Finally, the literature does not use realistic visual stimuli to feed the simulations, complicating the understanding of the impact of neural cyberattacks on real-world visual processes. 

\begin{table}[!htb]
\centering
\label{tableNeuralCyberattacks}
\resizebox{\columnwidth}{!}{
\begin{tabular}{@{}lllll@{}}
\toprule
\textbf{Work/Reference}& \textbf{Attacks}& \textbf{Taxonomy}& \textbf{Metrics used} & \textbf{Main results}\\
\midrule
López Bernal et al. \cite{Lopez_Bernal:cyberattacks_implants:2020} (2020) & FLO, SCA & From CNN (276 neurons) & \begin{tabular}{@{}l@{}} Number of spikes,\\ percentage of shifts, \\ dispersion of spikes \\ (temporal, spikes) \end{tabular} & \begin{tabular}{@{}l@{}} SCA $\downarrow$ number of spikes, $\uparrow$ dispersion of spikes, \\  FLO $\uparrow$ shift percentages, $\uparrow$ dispersion of spikes \end{tabular} \\ \midrule
López Bernal et al. \cite{Lopez_Bernal:jamming:2022} (2022) & JAM, FLO & From CNN (276 neurons) & \begin{tabular}{@{}l@{}} Number of spikes,\\ temporal dispersion\end{tabular} & \begin{tabular}{@{}l@{}} JAM $\downarrow$ number of spikes, $\downarrow$ temporal dispersion, \\ FLO $\uparrow$ number of spikes, $\uparrow$ temporal dispersion\end{tabular} \\ \midrule
López Bernal et al. \cite{LopezBernal:taxonomy_attacks:2023} (2023) & \begin{tabular}{@{}l@{}}FLO, JAM, SCA,\\ FOR, SPO, SYB,\\ SIN, NON\end{tabular} & From CNN (276 neurons) & Number of spikes & \begin{tabular}{@{}l@{}} JAM, NON $\downarrow$ number of spikes (short term),\\ SCA, NON $\downarrow$ number of spikes (long term) \end{tabular} \\ \midrule
López Madejska et al. \cite{Lopez_Madejska:realitic_topology:2024}(2024) & FLO, JAM & \begin{tabular}{@{}l@{}}Mouse's L4 from V1 using \\ point-neurons (450 neruons)\end{tabular} & Number of spikes & \begin{tabular}{@{}l@{}} FLO $\uparrow$ number of spikes,\\ JAM $\downarrow$ number of spikes \end{tabular}\\ 
\bottomrule
\end{tabular}}
\caption{Summary of works related to neural cyberattacks, presenting the attacks, the taxonomy, the metrics used, and the description of the most relevant results.}
\end{table}

\section{Proposed solution}
\label{sec:proposedSolution}
This section presents the design and implementation of the proposed solution. Particularly, this section describes the realistic neuronal topology used, the main components to build and simulate it using realistic visual stimuli, the neuronal cyberattacks implemented, and how the data obtained are analyzed.

\subsection{Solution design}

\begin{figure}[!htb]
\centerline{\includegraphics[width=0.4\columnwidth]{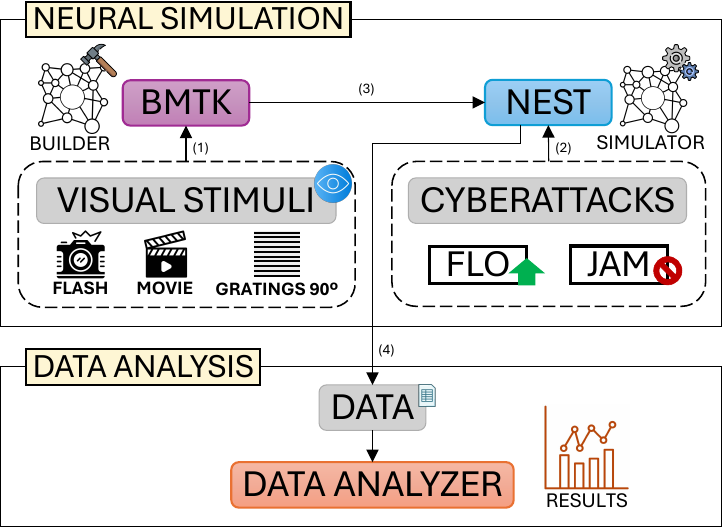}}
\caption{Design of the solution used to validate the impact of neural cyberattacks, indicating the main components of the solution and the data provided to each component.}
\label{designsolution}
\end{figure}

This section presents the solution proposed to validate the impact of two neuronal cyberattacks, FLO and JAM, in a realistic neuronal topology. These attacks were selected because they are conceptually simple and have low complexity in implementation and execution\addtxt{, unlike other neural cyberattacks that require a more challenging implementation}. Additionally, they allow the stimulation and inhibition of neural activity, respectively, covering both attacking approaches. \addtxt{They also differ in terms of duration and temporal resolution: one produces an instantaneous effect, while the other operates over a defined time window. This contrast allows the evaluation of different attack propagation mechanisms.} \figurename~\ref{designsolution} shows an overview of the proposed solution. First, the main component for building the topology is BMTK, which receives different stimuli as input offered by the authors of the realistic topology (1). The topology used in this work is proposed by Billeh et al. \cite{billeh:topology:2020}, who provide a complete reconstruction of a microcircuit from the primary visual cortex (V1) of mice. It comprises over 230,000 neurons with different populations and behaviors, including both excitatory and inhibitory neurons, modeling the characteristics of many morpho-electrical neuronal behaviors. \addtxt{This topology was chosen due to its high biological realism, as it not only integrates detailed morpho-electrical properties and synaptic connectivity, but also reconstructs the full six-layer architecture of mice V1. Compared to other available models, it offers the most comprehensive representation of visual processing, capturing both structural and functional features of the visual cortical microcircuit with a high degree of precision.} Just like the topology from Arkhipov et al. \cite{Arkhipov2018} mentioned in Section \ref{sec:relatedwork}, the topology used in this work provides two levels of granularity. The first level can simulate the behavior of the neurons in detail, and the electric and chemical processes that occur inside and between neurons (biophisiological approach), whereas the second level models the neuronal behavior according to the neuronal membrane potential in each instant (point-neurons). This point-neuron level comprises specific neurons called Gated Leaky Integrate and Fire (GLIF) neurons \cite{Teeter2018:GLIF} that model neuronal behavior based on sets of differential equations of neuronal voltage over time. Both models must be created with the BMTK tool \cite{BMTK}, and simulated accordingly to the approach, using NEURON \cite{NEURON} for the biophysically detailed model or NEST \cite{Gewaltig:NEST} for the point-neuron alternative. This research selected a point-neuron approach, which requires less computational resources and provides a sufficient approximation for the requirements of the investigation. \addtxt{This selection was based on the high computational cost associated with biophysically detailed models. In contrast, the point-neuron approach provides a sufficiently realistic representation of neural behavior without requiring detailed structural information about neurons. Additionally, it enables faster simulations, making it an efficient solution. Furthermore, analyzing the impact of cyberattacks does not require a highly detailed biophysical model, as the point-neuron approach proves sufficient for observing and evaluating their effects.}

The second main component to highlight is NEST. As previously mentioned, this tool is responsible for simulating the topology and it is where the source code was modified to include both neuronal cyberattack behaviors (2). After BMTK builds the topology, it is simulated by NEST. Therefore, both tools work together during the simulation (3). Once the results of the simulation are obtained, NEST transfers them to the data analysis block, where the data analyzer component is responsible for interpreting the data and, based on different impact metrics, will generate appropriate plots to study the impact of the cyberattacks visually (4).

\subsection{Solution implementation}
\label{sec:implementationSolution}
This section presents the implementation details conducted in this research for each main component representing the solution previously presented in \figurename~\ref{designsolution}. The following subsections explain relevant aspects of the tools employed and the impact metrics used to analyze the results obtained. The implemented source code and the dataset generated are publicly available in \cite{github-repo}.

\subsubsection{Implementation related to BMTK}

\begin{figure}[!htb]
\centerline{\includegraphics[width=0.5\columnwidth]{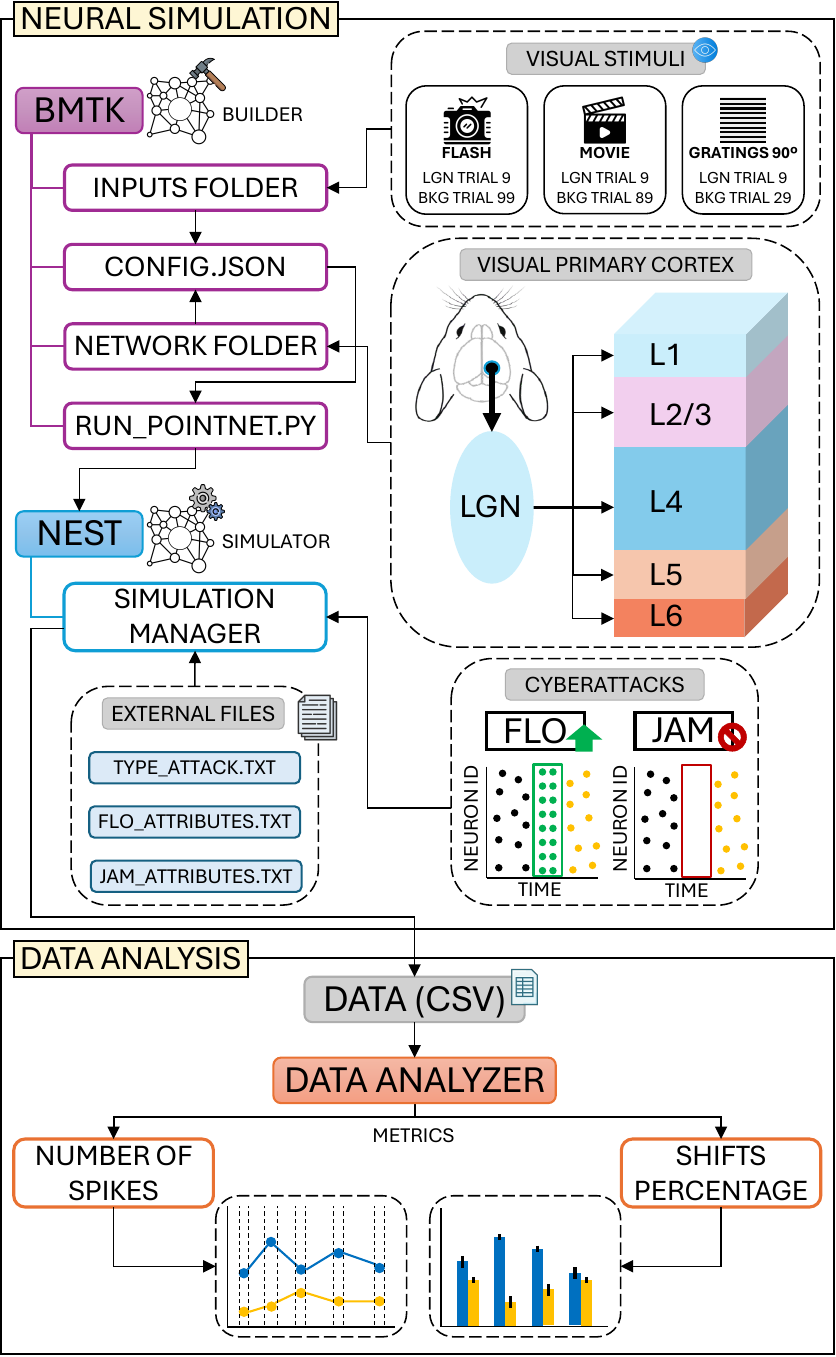}}
\caption{Simplified solution implementation conducted in this research, indicating the visual stimuli and network used, the main files modified to build and simulate the topology, the cyberattacks implemented, and the main metrics employed to generate plots to study the impact caused.}
\label{implementationsolution}
\end{figure}

\figurename~\ref{implementationsolution} presents the details of the components to build and simulate the topology selected and how the results are analyzed. The goal within BMTK is to build the complete topology and use different visual stimuli as input. To achieve it, it is relevant to consider firstly that the simulator has an \textit{inputs directory} containing files that represent the visual stimuli employed. These files include three kinds of stimuli: 
\begin{itemize}
\item \textbf{Light flash}: The flash effect lasts 2,500 ms and is composed of white, gray, and black events. The first represents a bright effect, the next describes a neutral visual scenario, and the last corresponds to a darkness situation. The flash stimulus starts with a gray screen for 500 ms, followed by 250 ms of a white visual stimulus (ON-flash); after that, the gray screen is presented again for 1,000 ms and, then, the black stimulus appears for 250 ms (OFF-flash). Finally, in the last 500 ms, the gray stimulus is presented again.
\item \textbf{Movie stimulus}: This effect is formed by 500 ms of the gray screen, and the 2,500 ms remaining correspond to three scenes (scenes 41, 42, and 43) from a 120-second clip of the opening of the film Touch of Evil (Zugsmith \& Welles, 1958), where each scene lasts one second. However, the last scene appears half a second in the simulation.
\item \textbf{Drifting gratings stimulus}: The drifting gratings are similar to the movie stimulus; it has 500 ms of gray screen, and the following 2,500 ms correspond to the stimulus. In this case, between the eight orientations that are available to choose as stimulus, the orientation chosen is 90 degrees (horizontal black and white lines) with a temporal frequency of 2Hz because these parameters have been taken by Billeh et al. as reference \cite{billeh:topology:2020}.
\end{itemize}

These three stimuli and the duration for each event are visually represented in \figurename~\ref{InstantScenario}, indicating the beginning and the end for each event within a stimuli. Nevertheless, this figure additionally contains information about the attacking instants, which will be described in detail in Section \ref{sec:experimentationscenario}. In addition to the visual stimuli previously mentioned, the simulation receives inputs from external brain regions, which act like background inputs (BKG). This activity follows a Poisson distribution at 1KHz.

\begin{figure}[!htb]
\centerline{\includegraphics[width=0.8\columnwidth]{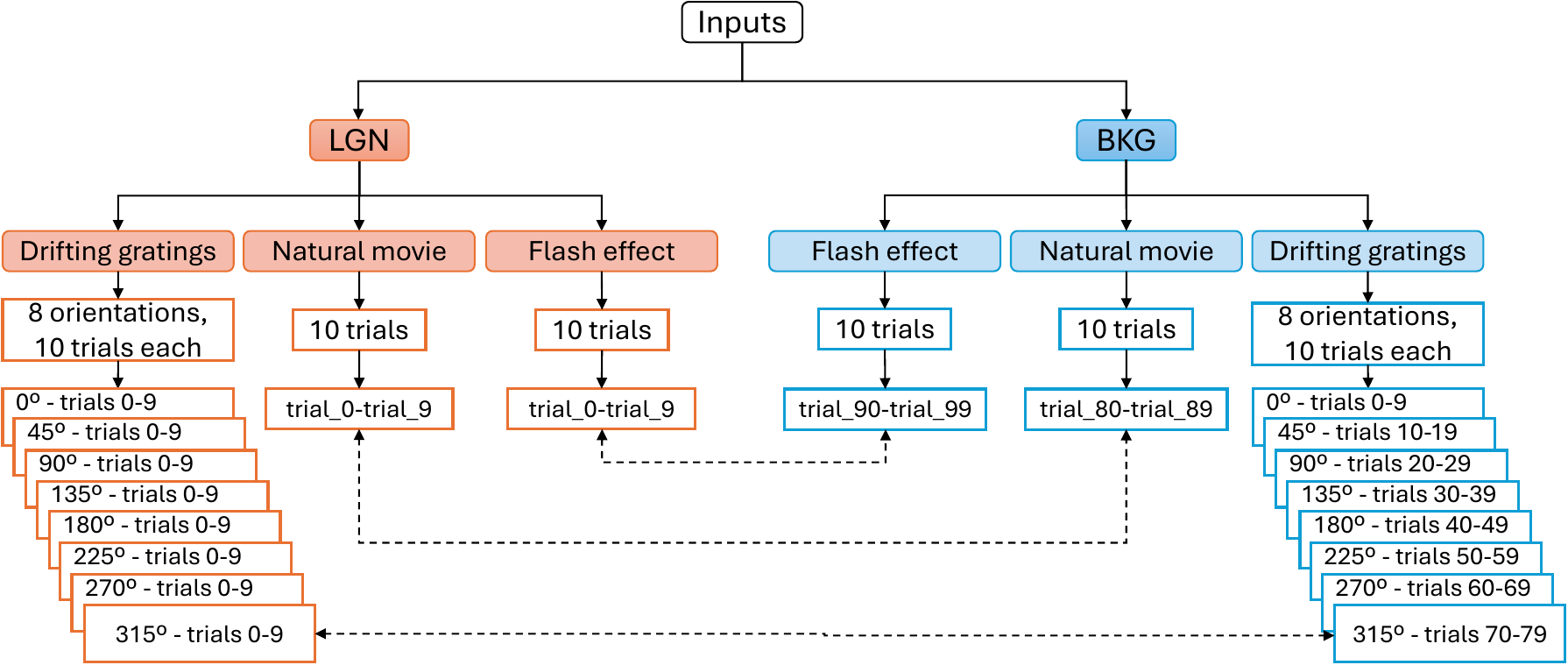}}
\caption{Graphical representation containing the distribution of trials per type of inputs, presenting the relationship between LGN and BKG trials.}
\label{Trials}
\end{figure}

\figurename~\ref{Trials} represents how the LGN and BKG stimulus provided by Billeh et al. are related. Focusing on the visual inputs represented within the LGN, Billeh et al. \cite{billeh:topology:2020} provide ten trials for each category of stimulus: ten for flash, another ten for the movie, and ten trials for each orientation of drifting gratings (80 trials in total). These trials are numerically labeled (trial 0 to trial 9) in all stimuli; even for each drifting gratings stimulus, each orientation has the same numbering. On the contrary, the BKG inputs represent background neural activity, which comprises 100 trials; unlike the LGN inputs, these trials are labeled from trial 0 to trial 99. According to this classification, trials 0 to 79 correspond to the drifting gratings trials from LGN, where each set of ten consecutive trials corresponds to an orientation. After that, trials 80 to 89 correspond to the movie stimuli and, finally, trials 90 to 99 are related to the flash effect. 

The second directory, as represented in \figurename~\ref{implementationsolution}, corresponds to the \textit{network folder}, which includes files related to V1, LGN, and BKG nodes and their connections. These files and those from the input directory are indicated in the \textit{config.json} file, which is the configuration file in charge of running the network. In this research, the \textit{config.json} file references to files for the complete topology, the stimulus trials for both LGN and BKG, the duration and resolution of the simulation, and the output directory. Then, this file is used by the \textit{run\_pointnet.py} script, which is responsible for building the topology and running the network. In this case, once the network is built and simulated according to the configuration file, it was necessary to change the output format within the \textit{run\_pointnet.py} script to facilitate the data analysis process by adding, for example, the cyberattack used, the attack instant and the LGN and BKG trials used.  

\subsubsection{Modifications done to NEST}

After building the model, the Python script from BMTK calls NEST to simulate the model. NEST comprises several files, but the \textit{simulation manager} file is the most relevant for this solution since it is where the cyberattacks have been implemented. In particular, this file manages the logic of the simulation for each time instant, being the best candidate to perform the attacks. All the necessary modifications performed over the source code of NEST to implement both neural cyberattacks can be consulted in López Madejska et al. \cite{Lopez_Madejska:realitic_topology:2024}. At the behavioral level, as shown in \figurename~\ref{implementationsolution}, performing FLO causes an increase of neural activity in a specific instant, whereas, in JAM, the set of neurons under attack would not have spiked during the period of application. Furthermore, this work used three files to define the parameters of the attacks to configure both cyberattacks externally to the simulation, avoiding changing the source code after testing each configuration. These files are called \textit{type\_attack.txt}, \textit{FLO\_attributes.txt}, and \textit{JAM\_attributes.txt}. The first file indicates which cyberattack has to be executed, while the second and third files contain in which instant the attack starts, the targeted number of neurons, and the voltage that used to simulate the attack. 

\subsubsection{Data analysis based on impact metrics}

When the neural simulation ends, the resulting spikes are saved in two different formats, CSV and SONATA \cite{sonata}, being the latter proposed by the same authors of the topology used. This work used the CSV format to analyze the data obtained using Python and, more specifically, Pandas. This research used two metrics from the literature, the number of spikes and the percentage of shifts, to verify the impact of the cyberattacks using the spikes generated by the simulation \cite{Lopez_Bernal:cyberattacks_implants:2020}. The first metric quantifies the spikes generated during the attacks compared to the spontaneous activity, whereas the second studies the delay of a spike in time compared to the spontaneous behavior. The results obtained by both metrics are represented visually using the Seaborn library in Python.

\section{Experimentation scenarios}
\label{sec:experimentationscenario}
This section describes different experimental scenarios used to implement and validate the neural cyberattacks mentioned in Section~\ref{sec:proposedSolution}. In particular, this work analyzed the most promising attack instants for each stimulus, as well as the most suitable trials per stimulus. Furthermore, this research studied the impact of different parameters used configure the behavior of the attacks. It is relevant to highlight that the experiments have been validated by brain anatomy experts from the University of Murcia, highlighting the viability of these threats against the human brain.  

\subsection{Study of attack instants for each stimulus}
\label{sec:attackinstants}
\begin{figure}[!htb]
\subfigure[Attack instants for FLO and JAM over the flash effect.]{\includegraphics[width=0.5\columnwidth ]{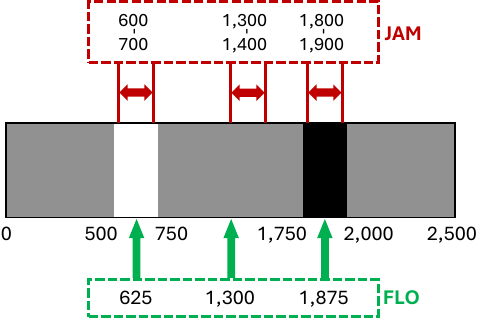}\label{InstantScenario/flash}}
\subfigure[Execution of FLO and JAM over the movie stimulus.]{\includegraphics[width=0.5\columnwidth ]{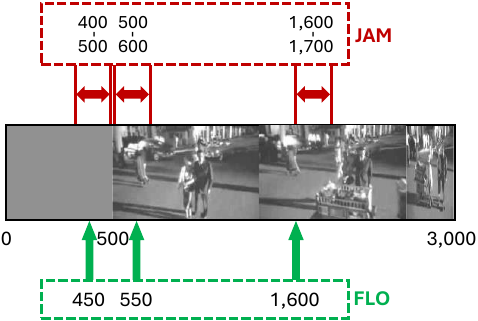}\label{InstantScenario/movie}}
\subfigure[Instants of attack for FLO and JAM over the drifting gratings stimulus.]{\includegraphics[width=0.5\columnwidth ]{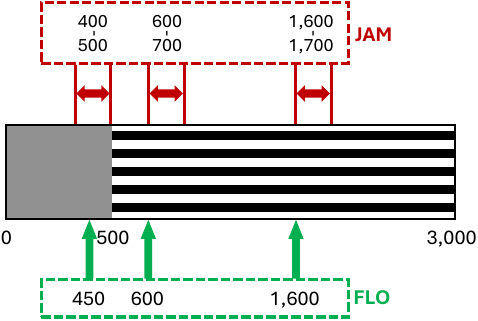}\label{InstantScenario/gratings}}
\caption{Visual representation of the three visual stimuli studied in this research, as well as the attack instants selected for FLO and JAM, represented with red and green colors, respectively.}
\label{InstantScenario}
\end{figure}

This section presents the attack instants selected for each cyberattack and each of the three stimuli tested. \figurename~\ref{InstantScenario} shows the instants selected for both attacks, differentiated by colors, for each stimulus and each relevant event within the stimulus. The selection was made based on the following criteria:

\begin{itemize}
    \item \textbf{Flash effect}: \figurename~\ref{InstantScenario/flash} shows which events are attacked within the flash effect. This work chose different situations that comprise the flash effect (white, gray, and black events) to analyze the impact of both cyberattacks. Starting with FLO, this attack is represented with a green color at the bottom of the image. It is important to mention that, although multiple attacks were tested over the same stimuli, the impact caused on each event was evaluated individually to avoid affecting the results of other executions. The first attack was performed over the ON-flash event, which comprises the time interval between the instants 500 ms and 750 ms. Then, observing the neural activity in this interval, there were a higher number of neural activations in the middle of the event. Based on that, this work opted for attacking at this moment, corresponding to the instant 625 ms. Meanwhile, the black event (OFF-flash) comprises the interval between 1,750 ms and 2,000 ms. Following the same approach as in the ON-flash stimulus, the attack instant was established to 1,875 ms. In the case of a neutral situation, there are three gray intervals in the flash stimulus, being the one in the middle the most promising since it has a greater duration and, consequently, allows evaluating the propagation of the cyberattack in a clearer way. Within this specific gray event, the objective is to attack when the influence of the white event over neural activity is over and before the black event starts. Therefore, this work selected the instant 1300 ms. In contrast, JAM is represented with a red color at the top of the image. Unlike the execution of FLO, JAM is applied over a time window. In order to study its evolution on each event previously indicated, this work selected an attack window of 100 ms, selecting the following intervals: 600-700 ms for the ON-flash, 1,800-1,900 ms for the OFF-flash, and 1,300-1,400 ms for the intermediate neutral event.
    \item \textbf{Movie stimulus}: This stimulus is represented in \figurename~\ref{InstantScenario/movie} and, as can be seen, this work has also selected three attack instants to maintain consistency across experiments. Focusing on FLO, the first attack instant selected is in the gray event placed before the movie, corresponding to the first 500 ms of the simulation. To evaluate the effect of FLO on neural activity before the beginning of the stimulus, the attack was executed in the instant 450 ms. The next instant under attack is right at the beginning of the stimulus, when there is a natural increase in neuronal spikes, specifically in the instant 550 ms. Finally, the last attack was performed in the middle of the stimulus (instant 1,600 ms) to analyze the effect on neural behavior after neural activity stabilizes. Furthermore, the JAM cyberattacks were executed during 100 ms, starting in the same time instants as in FLO. That is, one attack was performed in the neutral event (400-500 ms), the second at the beginning of the stimuli when there is a spontaneous peak in neural activity (500-600 ms) and, finally, in the middle of the movie (1,600-1,700 ms). 
    \item \textbf{Drifting gratings stimulus}: \figurename~\ref{InstantScenario/gratings} highlights that this stimulus also contains a gray event at the beginning, followed by the stimulus itself. Therefore, the first FLO attack instant is before the stimulus starts (instant 450 ms), as in the case of the movie. The next instant under attack is 600 ms, when the neurons have a peak of activity. The last instant corresponds to the middle of the stimuli and, more particularly, instant 1,600 ms, allowing to observe the effect of the attack over time. On the contrary, JAM is executed in the following temporal windows: 400-500 ms, 600-700 ms, and 1,600-1,700 ms, which start at similar instants as FLO and have a duration of 100 ms.
\end{itemize}

\subsection{Analysis of the trials selected}
\begin{figure}[!htb]
\centerline{\includegraphics[width=0.5\columnwidth]{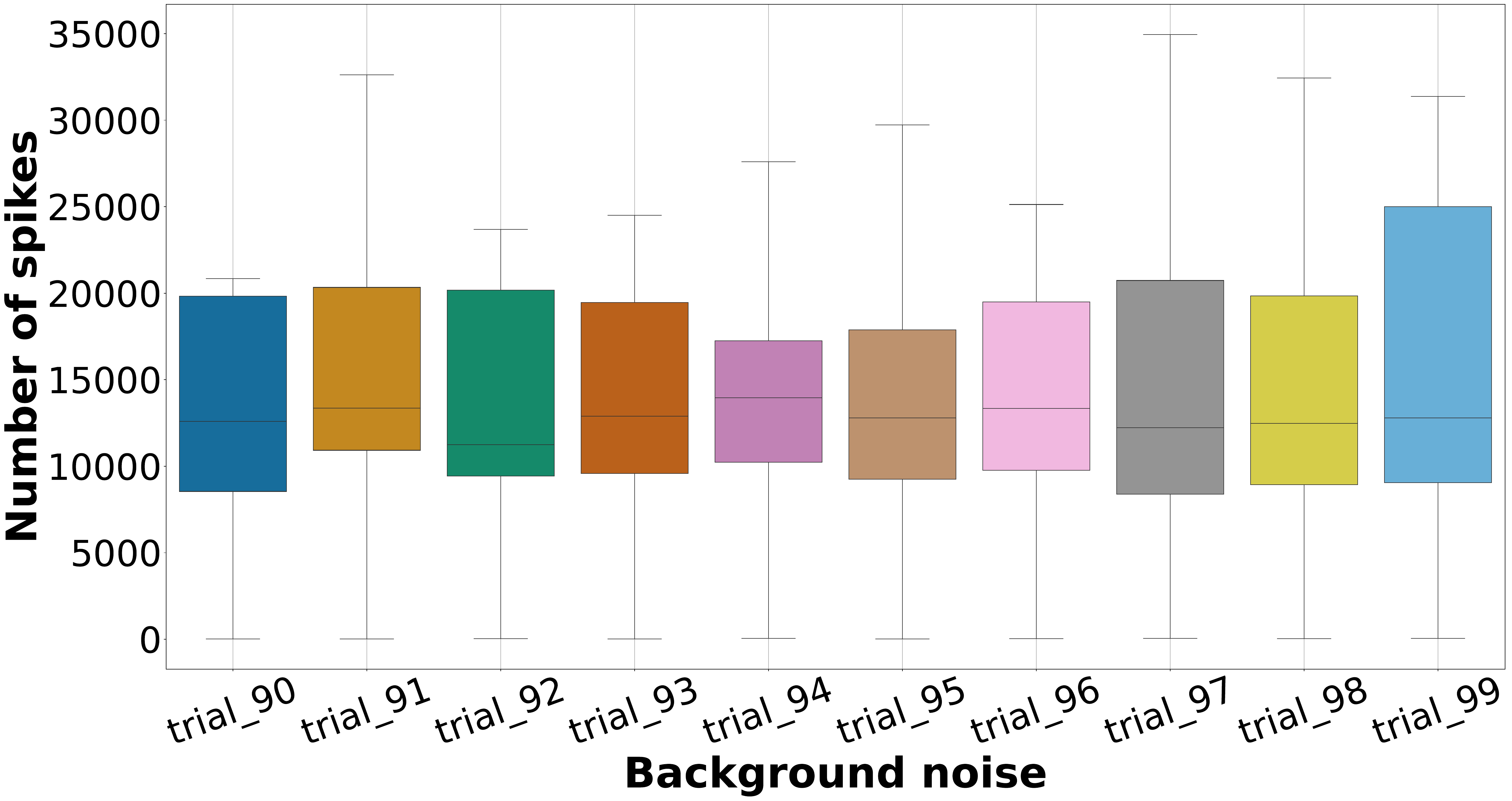}}
\caption{Evaluation of the number of spikes distribution for each BKG trial related to trial 9 from LGN, associated to the flash effect.}
\label{bkg}
\end{figure}

As specified in Section~\ref{sec:proposedSolution}, the topology used provides ten trials from LGN per stimulus. However, for the sake of simplicity in the experimentation, this work used only one LGN input, specifically trial 9, for all stimuli. This input was selected since Billeh et al. \cite{billeh:topology:2020} used it to illustrate their results on their paper, being representative of the behavior of the brain. Moreover, this LGN input has ten BKG trials associated for each stimuli. Nevertheless, the present work only requires one since its goal is to study the behavior and impact of the attacks on a base configuration, not being relevant to compare their effect on different trials. However, it is essential to determine which BKG input is the most suitable for the research. Thus, \figurename~\ref{bkg} presents the study of the distribution of spikes for each BKG input (trials from 90 to 99) associated to trial 9 of the flash stimulus to assess if there is relevant variability between combinations or if, on the contrary, they are very similar and, thus, any association would be adequate. The median distribution of the number of spikes metric changes between trials, but all oscillate within a specific range (10,000-20,000 spikes). Based on this, the BKG trial 99 was chosen because it presents a larger distribution of spikes. Moreover, Billeh et al. used this trial with trial 9 of the flash effect to describe their results. Besides, this research used trials highlighted by Billeh et al. for movie and drifting gratings: trial 89 for movie and trial 29 for drifting gratings.

\subsection{Attack parameters}
The cyberattacks were configured taking into consideration different parameters to analyze the effectiveness of the attacks. These parameters have been already explored in the literature by López Bernal et al. \cite{LopezBernal:taxonomy_attacks:2023, Lopez_Bernal:cyberattacks_implants:2020, Lopez_Bernal:jamming:2022} and López Madejska et al. \cite{Lopez_Madejska:realitic_topology:2024}. \addtxt{One of the key aspects of the configuration is that the cyberattacks are applied across all six layers of the V1 model. This decision ensures that the impact of the attacks is distributed throughout the entire cortical structure, rather than being limited to a specific region or subset of neurons. By targeting the entire visual cortex, the evaluation provides a more comprehensive understanding of the global effects induced by each attack.} This research considers as a parameter the number of neurons randomly attacked as specified in the literature, specifically selecting half (115,462) and a quarter (57,731) of the total number of neurons of the topology. Despite these parameters representing high percentages of the total number of neurons, these subsets are small compared to the whole of V1. Furthermore, each combination of attack parameters was executed ten times with the aim of analyzing if there existed a significant difference between distinct sets of randomly selected neurons. Additionally, the simulation duration is fixed at three seconds (3,000 ms) with a resolution of 0.25 ms (for evaluations per second), enough to assess the impact and propagation of the cyberattacks based on previous research \cite{LopezBernal:taxonomy_attacks:2023}.

Lastly, in FLO, this research establishes the voltage of the affected neurons to their individual threshold voltage (V\_th). This specific voltage forces each neuron to generate a spike\addtxt{, thereby ensuring the effectiveness of the attack. This voltage provokes the neuron} \rmvtxt{sufficient activity} to transmit an impulse \change{to connected neurons}{to those neurons with which it has synapses}. In JAM, the voltage of the neurons is forced to its minimum possible value\addtxt{, corresponding to the reset potential} (V\_reset) to prevent the targeted neurons from having activity during the attack window. \addtxt{This abrupt reduction effectively maintains the neurons below their firing threshold, thereby suppressing their ability to generate an action potential.} These parameters are selected after experiments performed by the literature \cite{Lopez_Bernal:cyberattacks_implants:2020} \cite{Lopez_Madejska:realitic_topology:2024}, where these works indicated that the ideal voltages to cause a greater effect are the values close to the threshold voltage in the case of FLO and the values near the minimum voltage in JAM. Furthermore, each combination of parameters was executed ten times to offer enough variability of randomly-selected neurons to obtain valid conclusions. Finally, \tablename~\ref{tableAttacksConfiguration} summarizes the parameters selected in this research for each cyberattack used.

\begin{table}[!htb]
\centering
\label{tableAttacksConfiguration}
\resizebox{\columnwidth}{!}{
\begin{tabular}{@{}lllllll@{}}
\midrule
\textbf{Attack}&  \textbf{Visual stimuli}& \textbf{Background}& \textbf{Attack instants} & \textbf{Targeted neurons} & \textbf{Voltage used} \\ \midrule
FLO & Flash & Trial 99& 625 ms, 1,300 ms, 1,875 ms& 25\%, 50\% & V\_th\\ \midrule
JAM & Flash & Trial 99& 600-700 ms, 1,300-1,400 ms, 1,800-1,900 ms& 25\%, 50\% & V\_reset\\ \midrule

FLO & Movie & Trial 89& 450 ms, 550 ms, 1,600 ms& 25\%, 50\% & V\_th\\ \midrule
JAM & Movie & Trial 89& 400-500 ms, 500-600 ms, 1,600-1,700 ms & 25\%, 50\% & V\_reset\\ \midrule

FLO & Drifting gratings 90º & Trial 29& 450 ms, 600 ms, 1,600 ms& 25\%, 50\% & V\_th\\ \midrule
JAM & Drifting gratings 90º & Trial 29& 400-500 ms, 600-700 ms, 1,600-1,700 ms& 25\%, 50\% & V\_reset\\

\bottomrule
\end{tabular}}
\caption{Implemented configurations for each neural cyberattack executed ten times, using trial 9 form LGN for all stimuli.}
\end{table}

\section{Analysis of the results obtained}
\label{sec:results}

\begin{figure}[!htb]
\centerline{\includegraphics[width=0.5\columnwidth]{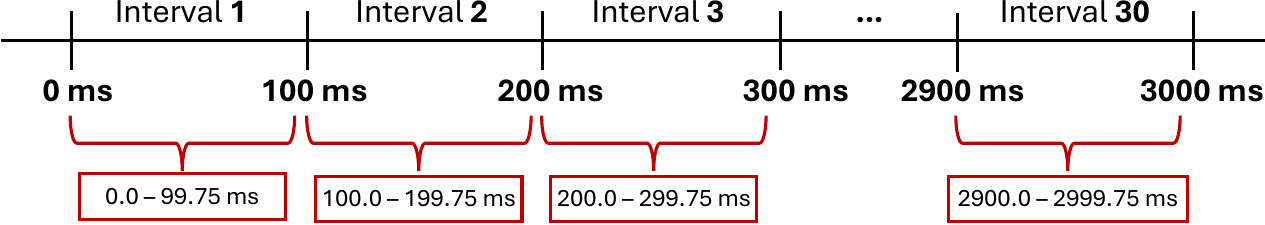}}
\caption{Temporal division of the simulation in intervals, indicating the temporal ranges covered per interval, and highlighting the first three intervals and the last one.}
\label{Intervals}
\end{figure}

This section presents the analysis of the impact of each cyberattack over the three stimuli selected in this manuscript. The impact of the attacks is verified using two metrics from the literature \cite{Lopez_Bernal:cyberattacks_implants:2020}, the number of spikes and the percentage of shifts, previously mentioned in Section \ref{sec:implementationSolution}. As the simulation has a duration of 3,000 ms and, for each millisecond, there are four values represented as spikes due to the simulation resolution (0.25), this work simplifies the visualization of the results by aggregating the information in temporal intervals of 100 ms, resulting in 30 intervals. Figure \ref{Intervals} illustrates the temporal ranges of each interval, specifically the first three intervals and the last one for simplicity. For example, the third interval comprises the slice of the simulation between the instants 200.0 and 299.75. Therefore, if the attack is executed, for example, in the instant 200 ms, the impact will manifest in the third interval.

\subsection{Impact study over the flash effect}

\begin{figure}[!htb]
\subfigure[\change{Impact on the number of spikes over 25\% of the neurons.}{Impact on the number of spikes over 25\% (left figure) and 50\% (right figure) of the neurons. The red numbers indicated the three events under attack individually, specified in Section~\ref{sec:attackinstants}.}]{\includegraphics[width=\textwidth]{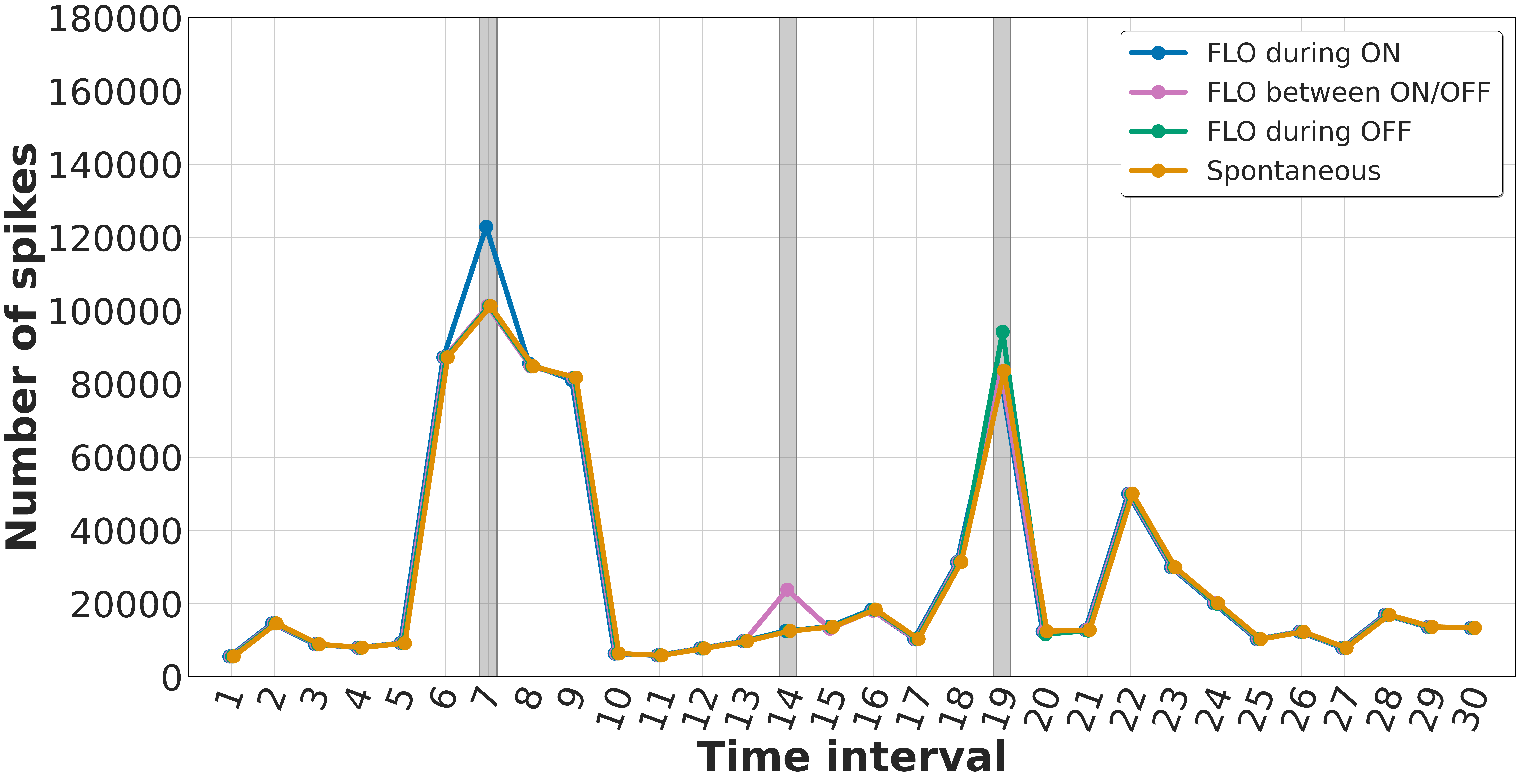}\label{FLOFlash/a}}
\subfigure[\change{Impact on the percentage of shifts over 50\% of the neurons.}{Impact on the percentage of shifts over 25\% (left figure) and 50\% (right figure) of the neurons. Events attacked individually are indicated with red numbers.}]{\includegraphics[width=\textwidth]{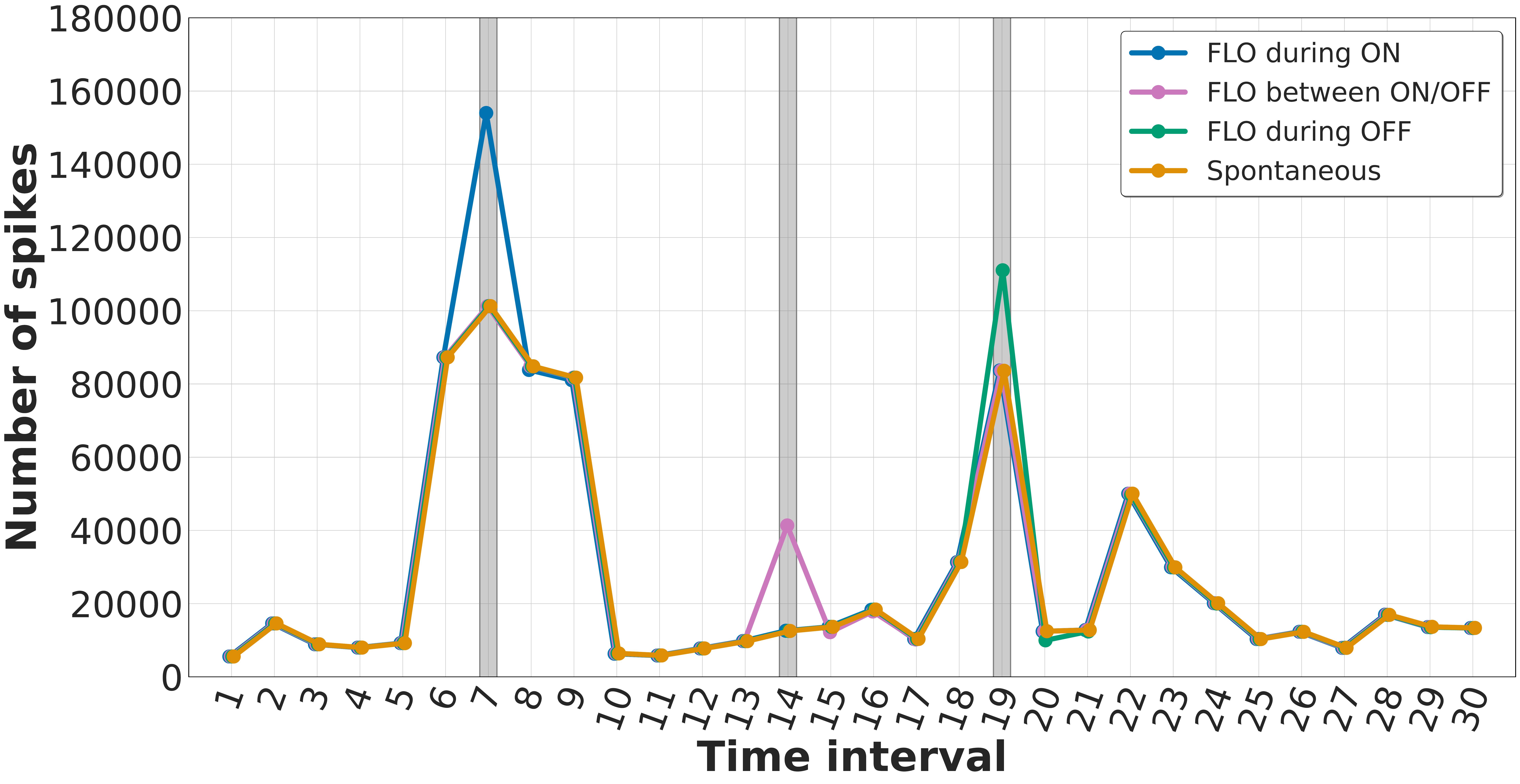}\label{FLOFlash/b}}
\caption{Analysis of the impact caused by FLO over the flash event considering the number of spikes and shift percentage metrics when attacking three different events, differentiated by attacking 25\% and 50\% of the total number of neurons. The random selection of neurons has been performed ten times per experiment to analyze the variability.}
\label{FLO Flash}
\end{figure}

Starting with FLO, \figurename~\ref{FLO Flash} shows the behavior of the cyberattack over the flash effect, differentiating by the number of neurons targeted. The attack intervals are represented in the plots with a gray shade \addtxt{and red numbers} to visualize the duration of the cyberattack. The plots containing the attack over 25\% of the neurons are represented on the left side \rmvtxt{(~\ref{FLOFlash/a} and ~\ref{FLOFlash/c})}, whereas the plots with 50\% of the neurons under attack are on the right side \rmvtxt{(~\ref{FLOFlash/b} and ~\ref{FLOFlash/d})}. It is important to highlight that this distribution is maintained in all subsequent figures representing the results obtained. As observed in the plots, the attack instants selected (see Section \ref{sec:attackinstants}) are distinguished by blue (ON-flash), pink (neutral event), and green (OFF-flash) lines in the case of the number of spikes metric, and bars for the shifts percentage metric. Besides, in the number of spikes metric, the lines are compared with the spontaneous behavior (orange line). Moreover, each line or bar contains the results for the ten executions performed per combination of attacking parameters, showing the average result per time interval. 

\figurename~\ref{FLOFlash/a}\addtxt{, the left plot,} depicts the number of spikes analysis when the attack affects 25\% of the total neurons. First, this figure shows that, in the ON-flash interval, there is an increase of 21,667 spikes (\change{21,40\%}{21.40\%}) compared to spontaneous behavior. On the contrary, attacking during the gray event, neural activity augments in 11,284 spikes (89.97\%), whereas, in the OFF-flash, FLO has caused 10,651 \change{spikes more}{more spikes} (\change{12,74\%}{12.74\%}) over the spontaneous simulation. In contrast, \change{\ref{FLOFlash/b}}{the right figure} represents the number of spikes when 50\% of neurons are attacked. In the first event, the attack causes an increase of 52,742 spikes (\change{52,08\%}{52.08\%}), although the augmentation in the neutral event is 28,815 spikes (\change{229,7\%}{229.7\%}). Finally, in the OFF-flash, there have been 27,444 additional spikes (\change{32,83\%}{32.83\%}) over the spontaneous signaling. As can be seen, the gray event is the most impacted as neurons spontaneously exhibit a lower response than during other events; thus, any malicious overstimulation will cause a notorious variation. Nevertheless, it is important to highlight that the attack impact can vary depending on the stimulus presented since neurons naturally behave in different ways according to visual inputs. Moreover, after attacking each event, the impact regarding the number of spikes metric suffers a propagation that lasts around two or three intervals (200-300 ms) until the neurons return to normal behavior. The recovery of the spontaneous state occurs in all experiments as the topology is composed of realistic individual neurons, which tend to synchronize again with their normal activity based on the visual stimulus. 

Regarding the analysis of shifts percentages, \addtxt{the left side of} \figurename~\ref{FLOFlash/b} represents the results when the attacks affect 25\% of the neurons. First of all, attacking during the ON-flash causes a 61\% shift increase. However, attacking during the gray event forces 94\% of the spikes to be delayed, while the attack during the OFF-flash generates 16\% of shifts. In contrast, the percentage of shifts when 50\% of neurons are affected is represented on \change{~\ref{FLOFlash/d}}{the right side of \figurename~\ref{FLOFlash/b}}. Attacking during the ON-flash causes a variation of 71\%, whereas the neutral event and OFF-flash obtain 98\% and 29\% shifts, respectively. As observed, the gray event obtained the higher percentages as in the last metric. Nevertheless, there is a substantial increase in all intervals under attack, followed by fluctuations maintained until the end of the simulation (around 40\%-60\%). All these results indicate that, although neurons seem to return to the spontaneous activity according to the number of spikes metric, the attack causes a delay of the activations in time that gets propagated over the whole simulation, highlighted by the maintained effect observed by this metric. 

\begin{figure}[!htb]
\subfigure[\change{Number of spikes when JAM inhibits 25\% of the neurons.}{Number of spikes when JAM inhibits 25\% (left figure) and 50\% (right figure) of the neurons, on the three events indicated with gray shades and red numbers.}]{\includegraphics[width=\textwidth]{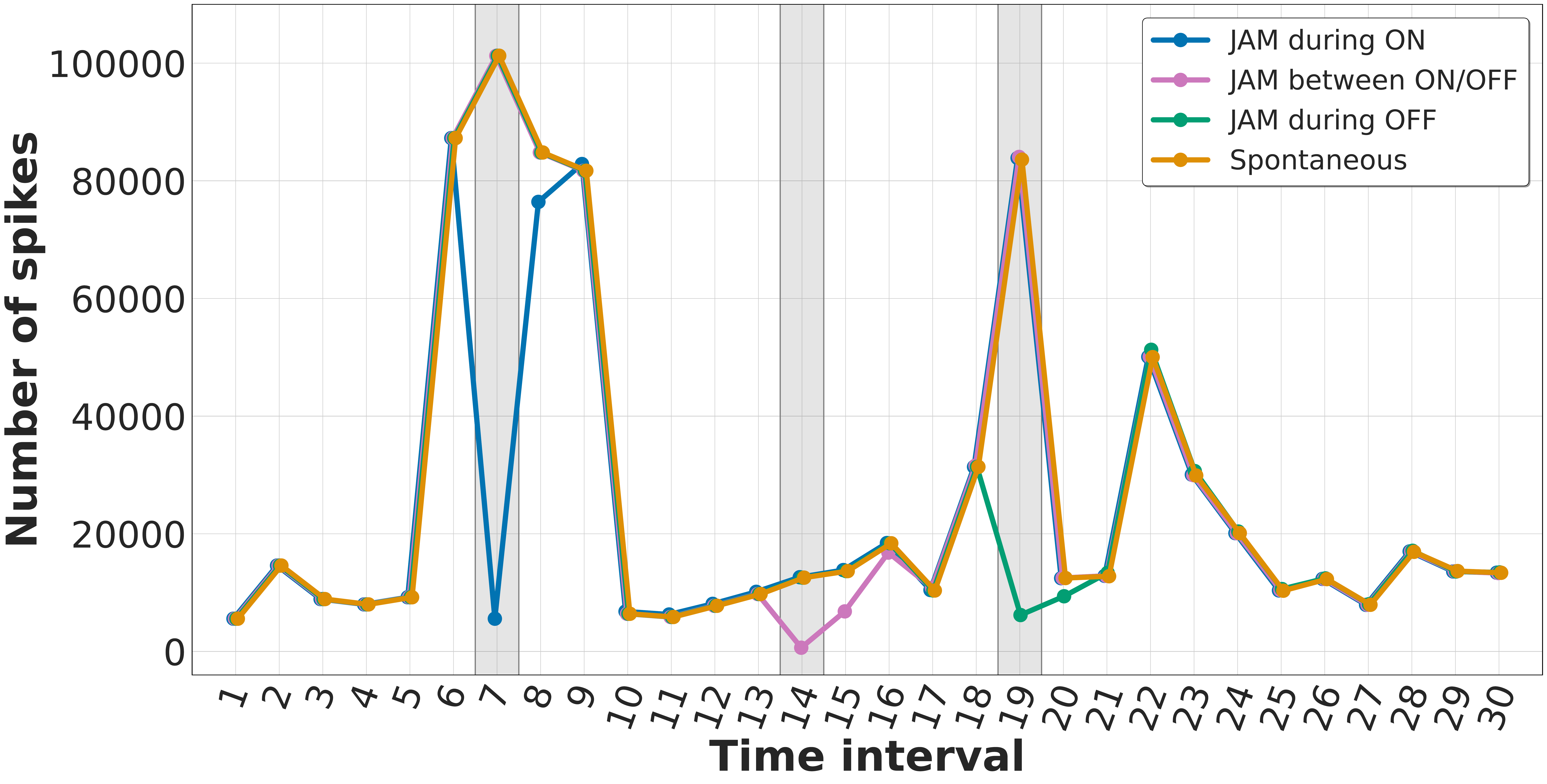}\label{JAMFlash/a}}
\subfigure[\change{Number of spikes when targeting 50\% of the neurons.}{Shifts percentages of the impact over 25\% (left figure) and 50\% (right figure) of the neurons. Red numbers and gray shades highlighted the events under attack.}]{\includegraphics[width=\textwidth]{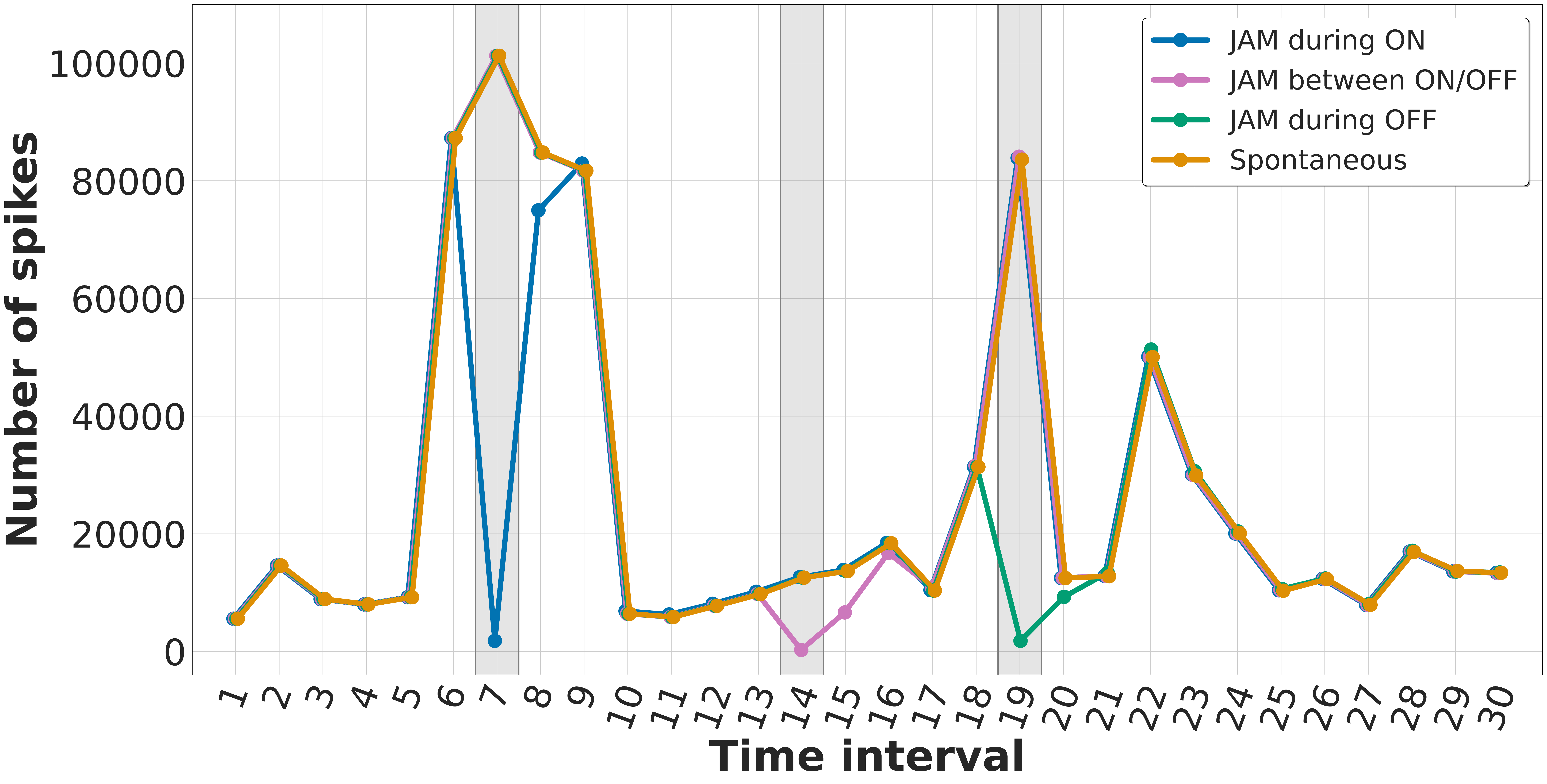}\label{JAMFlash/b}}
\caption{Study of the impact induced by JAM over the flash event, taking into consideration the number of spikes and shift percentage metrics. In particular, it details the impact over three different events when attacking 25\% and 50\% of the neurons, evaluating ten times each combination of parameters.}
\label{JAM Flash}
\end{figure}

\figurename~\ref{JAM Flash} represents the impact of JAM when inhibiting different time windows selected in Section \ref{sec:attackinstants}. Starting with \addtxt{the left plot on }\figurename~\ref{JAMFlash/a}, where the 25\% of the neurons are inhibited, the ON-flash presents a mean decrease of 95,696 spikes (-94.50\%). In contrast, attacking during the neutral interval reaches a decrement of 11,910 spikes (-94.96\%), while the impact for the OFF-flash decreases the number of spikes by 77,413 spikes (-92.60\%) compared to spontaneous behavior. Next, \change{\ref{JAMFlash/b}}{the right figure} shows that the attack produces a decrease of 99,461 spikes (-98.22\%) in the white event, while in the gray event the number of spikes decreases by 12,290 spikes (-97.99\%). Finally, the OFF-flash there have been 81,794 spikes less (-97.85\%). Once JAM has finished, the inhibited neurons generate a high peak of activations, caused by the liberation of the neurons after the attack, which tend to return to their normal behavior in a coordinated way. The results highlight the similarity between attacking both number of neurons, implying that it is not necessary to inhibit a large set of neurons to produce a significant impact. Depending on the event the attack affects, the propagation of the impact differs. Attacking the ON-flash and gray events takes around 11 intervals (1,100 ms) to converge. In contrast, in the OFF-flash, the neurons take seven intervals (700ms) to return to normal activity. 

In the percentage of shifts metric, represented by \figurename~\ref{JAMFlash/b}\addtxt{, the left plot} for attacking 25\% of the neurons, the attack interval of the ON-flash reaches 33\% of shifts, the gray event 16\%, and the OFF-flash 48\% of shifts. Oppositely, \change{\ref{JAMFlash/d}}{the right side of \figurename~\ref{JAMFlash/b}} depicts that the white event obtains a percentage of 12\%. In contrast, the neutral event reaches a percentage of 4\%, while the black event obtains 27\% of shifts. As can be observed, the reduction in the number of spikes directly affects the percentage of shifts since having a very low number of spikes per interval will restrict the variability of the spikes over time. Moreover, the high peak produced after the cyberattack, already explained in the previous metric, is also represented in this metric after the gray shades. More particularly, the intervals after the attacks start to fluctuate until the percentages stabilize between 50\%-70\%.

\subsection{Impact study over the movie stimulus}

\begin{figure}[!htb]
\subfigure[\change{Total number of spikes when FLO targets 25\% of the neurons.}{Total number of spikes when FLO targets 25\% (left image) and 50\% (right image) of the neurons. Events under attacks are indicated by red numbers and gray shades.}]{\includegraphics[width=\textwidth]{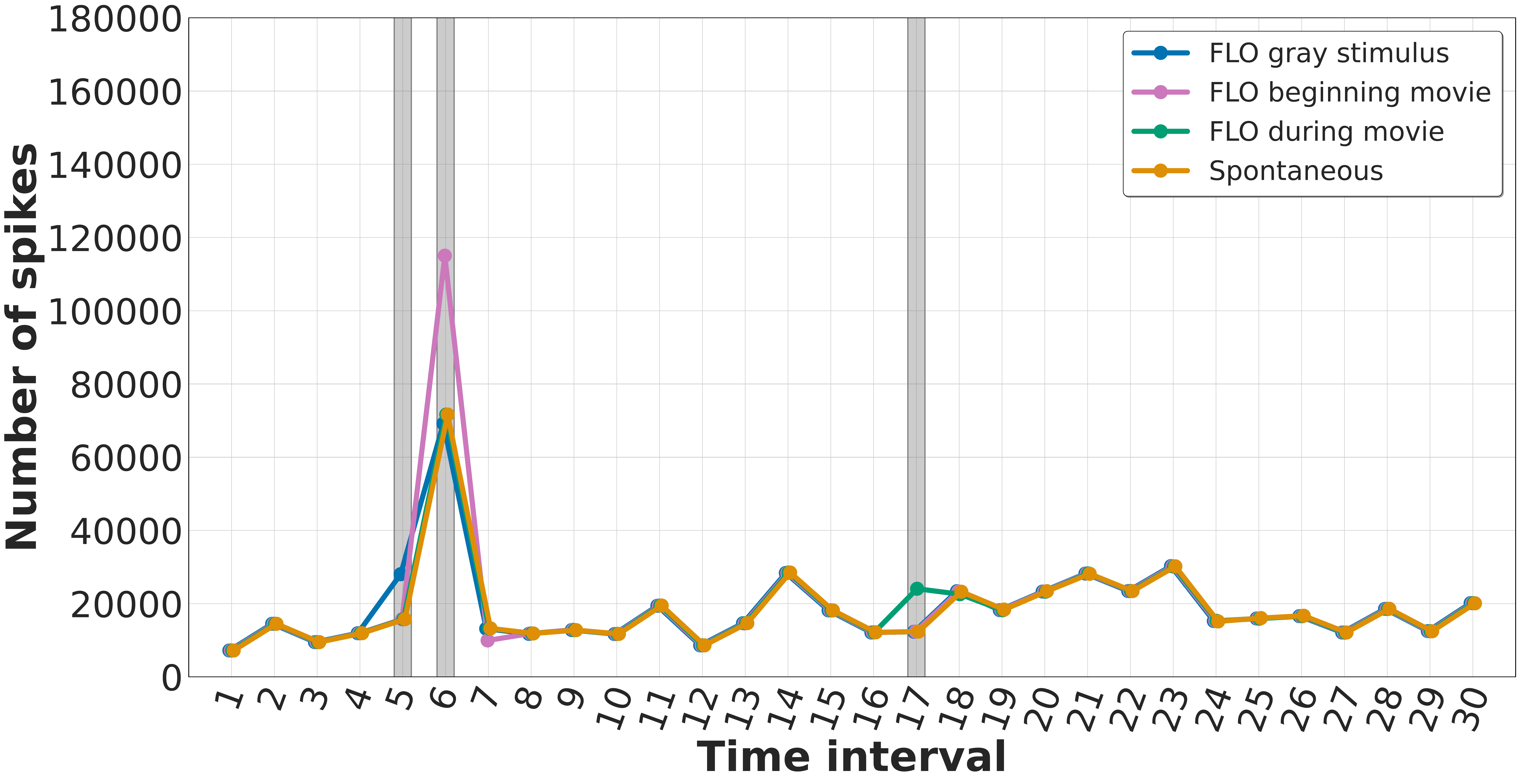}\label{FLOMovie/a}}
\subfigure[\change{Number of spikes analysis for the impact over 50\% of the neurons.}{Shifts percentages over 25\% (left image) and 50\% (right image) of the neurons. The red numbers indicated the three events under attack individually. }]{\includegraphics[width=\textwidth]{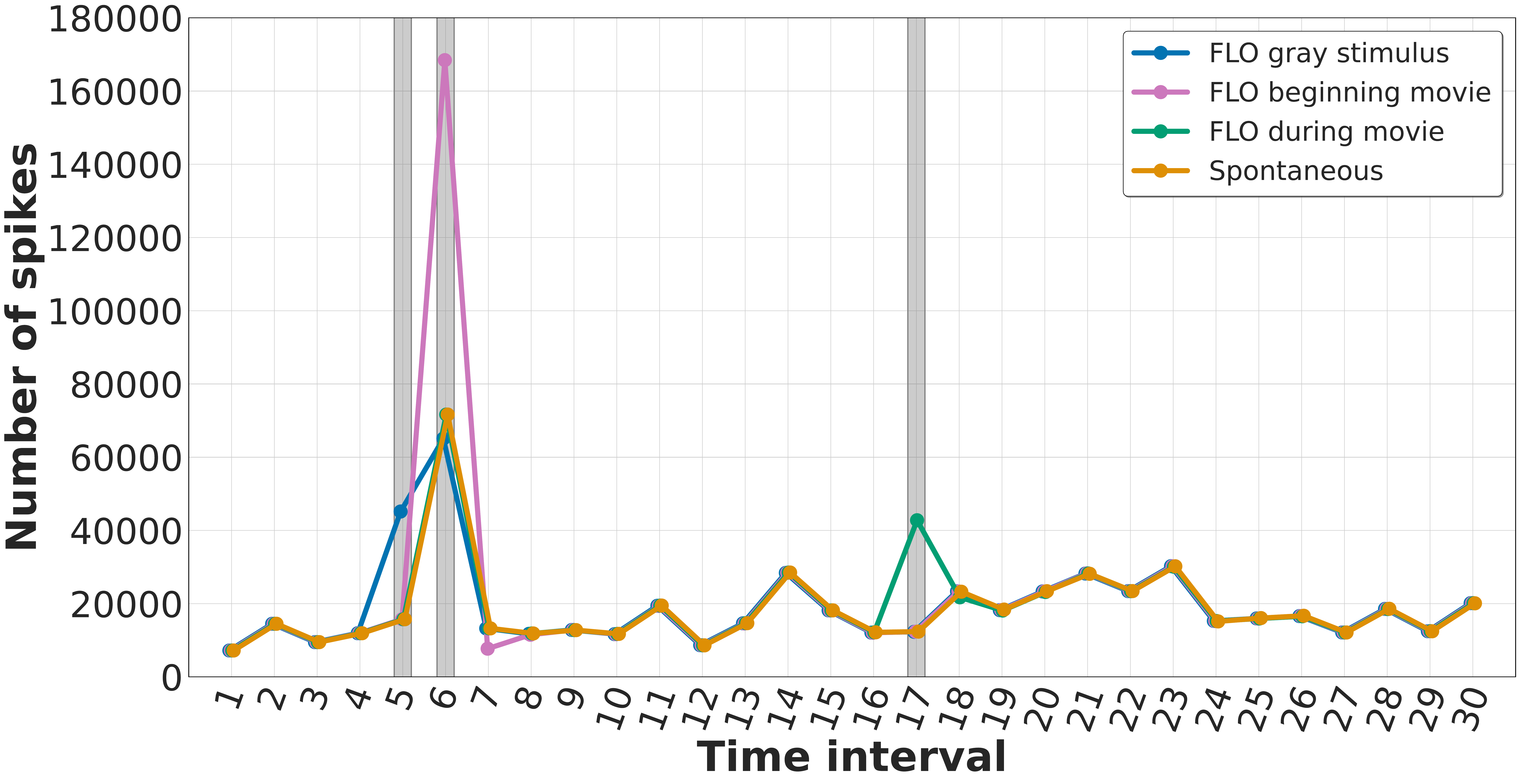}\label{FLOMovie/b}}
\caption{Analysis of the impact induced by FLO over the movie stimulus considering the number of spikes and shift percentage metrics when targeting three events. It also differentiates the impact caused by affecting 25\% and 50\% of the neurons from the topology. The variability presented in the figure corresponds to ten distinct executions for each combination of parameters, where each execution performs a random selection of neurons.}
\label{FLO Movie}
\end{figure}

\figurename~\ref{FLO Movie} shows the impact of FLO in the three attack instants selected.\addtxt{ The left plot of }\figurename~\ref{FLOMovie/a} presents the number of spikes metric when attacking 25\% of the neurons, where attacking during the gray event causes an increase of 12,297 spikes (\change{78,12\%}{78.12\%}). Then, when FLO was executed at the beginning of the movie (after the transition from the neutral event), the number of spikes had an increase of 43,394 spikes (\change{60,57\%}{60.57\%}). Finally, when FLO was executed in the middle of the movie, the attack caused an increase of 11,702 spikes (\change{94,64\%}{94.64\%}). After that, \change{~\ref{FLOMovie/b}}{the right figure} represents the results of the same three instants for 50\% of the neurons. First, in the neutral event, the number of spikes metric augments in 29,422 spikes (\change{186,92\%}{186.92\%}) according to spontaneous activity. Secondly, in the instant corresponding to the beginning of the movie stimulus, there was an augment of 96,798 spikes (\change{135,11\%}{135.11\%}). Finally, during the movie, FLO causes an increase of 30,410 spikes (\change{245,95\%}{245.95\%}) compared to spontaneous behavior. In both number of neurons, it is clear that when the neurons have a lower response to an event, any overstimulation generates a notorious increase in spikes. However, if only the percentage of increase is considered, the most affected event is the instant during the movie. Nevertheless, from the number of spikes perspective, the greater impact can be observed during the sixth instant, representing the start of the movie stimulus, in both sets of neurons. This last situation can be explained since neurons change drastically from reacting to a neutral event to a new and more complex event, causing an increase of spikes in comparison to other events. In addition, the recovery to normal neuronal activity in the three attack instants is around 200-300 ms.

Focusing on spike shifts, \figurename~\ref{FLOMovie/b} highlights\addtxt{,  in the plot that represents 25\% of the neurons attacked,} that the gray event gets delayed by 70\%, the second instant obtains a percentage of shifts of 75\% and, during the last instant, the attack causes a variation of 93\%. In contrast, \change{\ref{FLOMovie/d}}{the right side of \figurename~\ref{FLOMovie/b} }represents the shifts percentage when 50\% of the neurons are affected. The neutral event obtains 82\% of shifts then, the event at the beginning of the movie gets delayed by 84\% and, finally, during the movie stimulus, the percentage obtained is 98\%. As in the last metric, the most sensitive effect corresponds to the movie stimulus since the neurons generate spikes in response to an event where they do not normally react too strongly. As an overview, all attack instants present a large shifts percentage that tends to decrease over time after the attack. Even at the end of the simulation, this metric maintains high values between 40\%-50\%. 

\begin{figure}[!htb]
\subfigure[\change{Number of spikes of the impact over the 25\% of the neurons.}{Number of spikes of the impact over the 25\% (left image) and 50\% (right image) of the neurons. Attack instants are highlighted by red numbers.}]{\includegraphics[width=\textwidth]{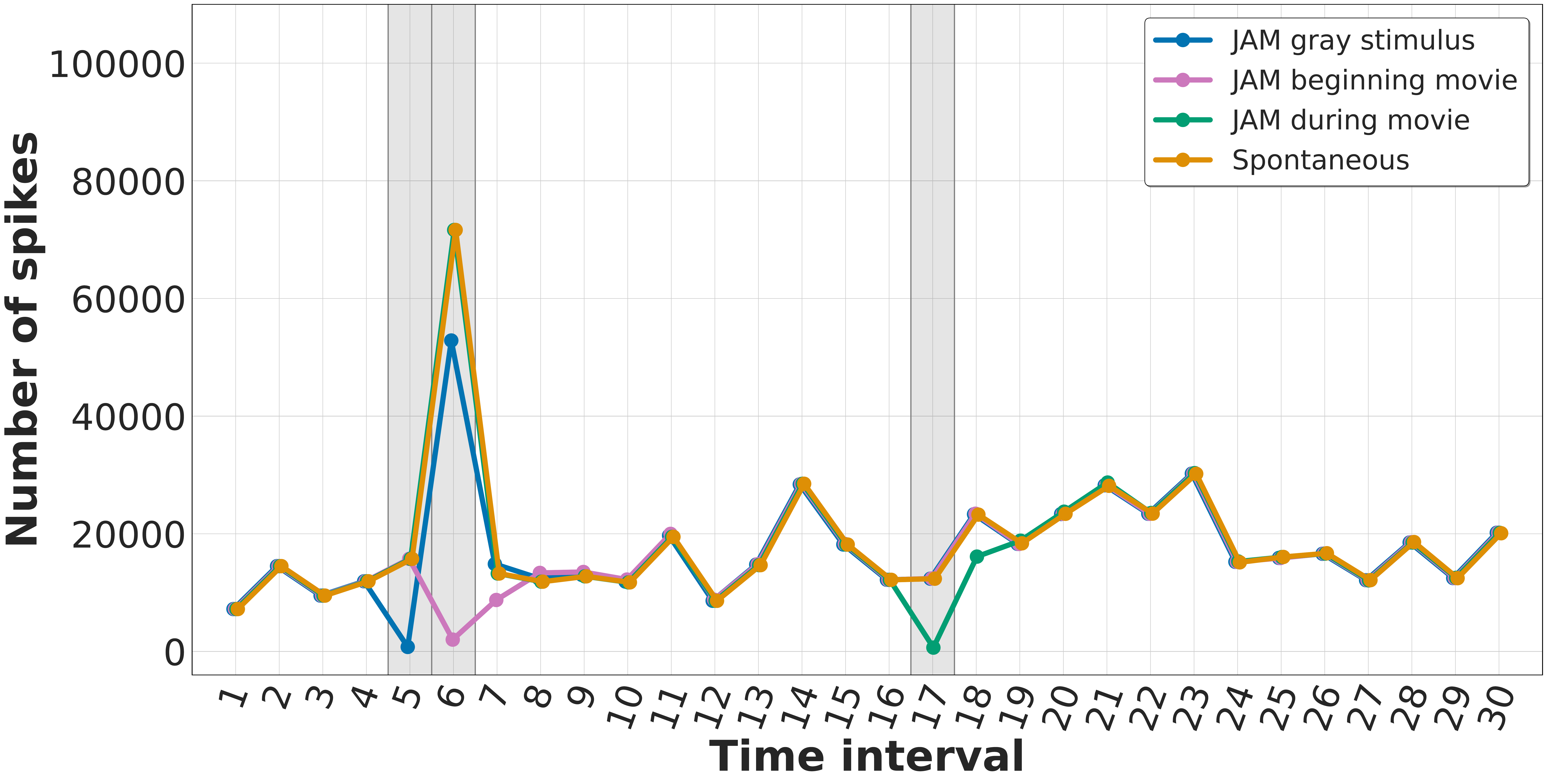}\label{JAMMovie/a}}
\subfigure[\change{Total number of spikes when JAM affects the 50\% of the neurons.}{Percentage of shifts of the attack impact, affecting 25\% (left image) and 50\% (right image) of the neurons. Events under attack are highlighted using red numbers and gray shading.}]{\includegraphics[width=\textwidth]{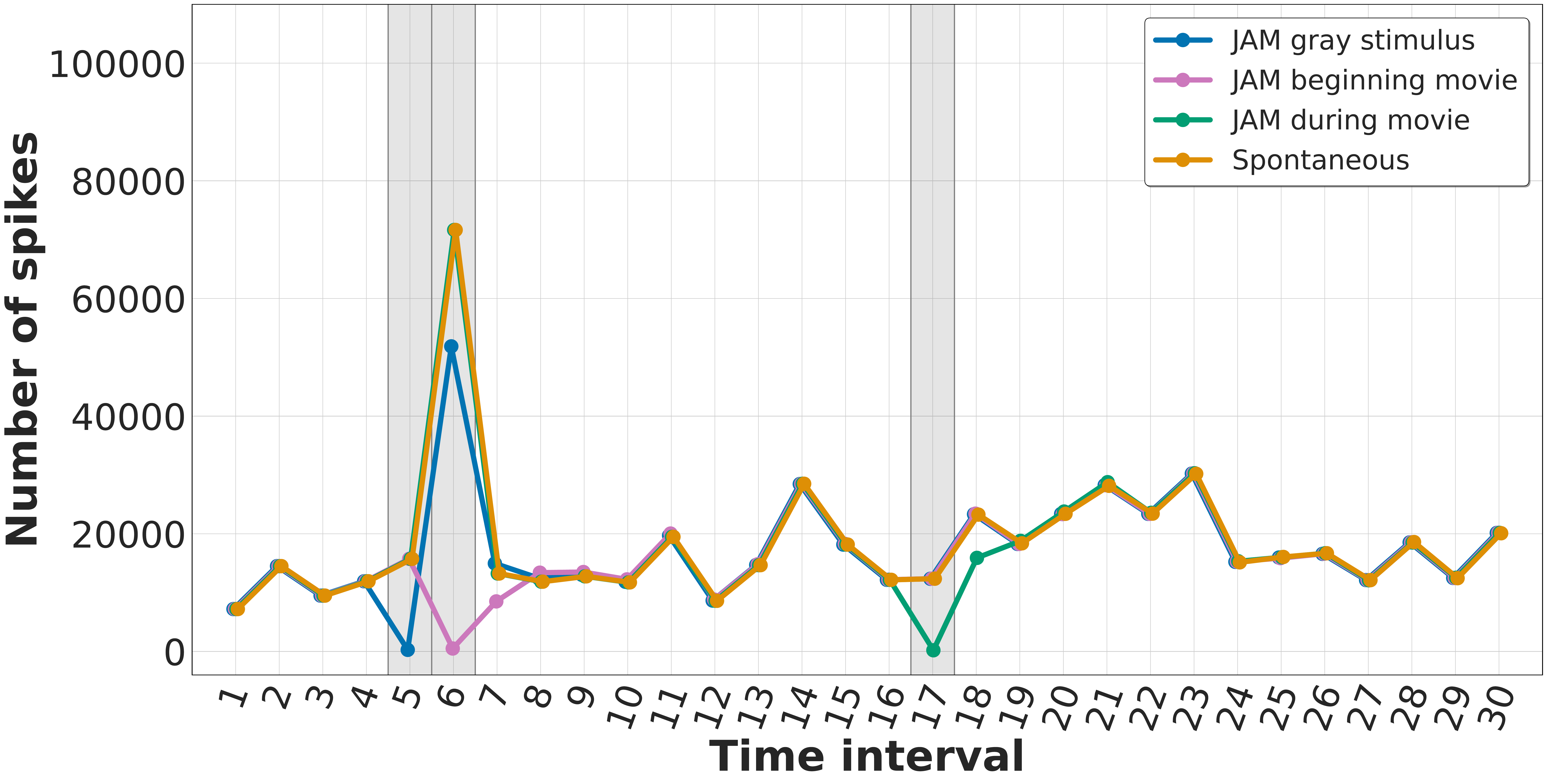}\label{JAMMovie/b}}
\caption{Study of the impact caused by JAM over the movie stimulus considering the number of spikes and shift percentage metrics differentiated by three different events and the 25\% and 50\% affected neurons. The random selection of neurons has been performed ten times per experiment to analyze the variability.}
\label{JAM Movie}
\end{figure}

\figurename~\ref{JAM Movie} depicts the impact of JAM over the previously mentioned events. On the one hand, \addtxt{the left plot of} \figurename~\ref{JAMMovie/a} presents the number of spikes analysis when JAM inhibits 25\% of the neurons. Attacking during the neutral event causes a decrease of 14,975 spikes (-95.14\%), whereas performing JAM on the second event causes a decrement of 69,636 spikes (-97.20\%). The interval corresponding to the middle of the movie stimulus falls to 11,714 spikes (-94.74\%) when performing the attack. On the other hand, \change{\ref{JAMMovie/b}}{the plot that represents 50\% of the neurons attacked on \figurename~\ref{JAMMovie/a}} shows the same analysis as the previous figure for 50\% of the neurons. The gray event suffers a decrease of 15,471 spikes (-98.29\%), then JAM generates a reduction of 71,147 spikes (-99.31\%) in the event representing the start of the movie, and the diminishment in the next event is 12,166 spikes (-98.40\%). In this case, the neurons take seven intervals (700 ms) to return to spontaneous behavior when executing the attack in the gray event. In contrast, in the attack intervals corresponding to the beginning and the middle of the movie, the neurons take six intervals (600 ms) to recover to normal activity. As can be seen, both numbers of neurons have similar results; therefore, as in the previous stimulus, it is not necessary to attack many neurons to produce a notorious impact. 

In the analysis of the shifts, and particularly attending to \addtxt{the left plot of} \figurename~\ref{JAMMovie/b}, the gray event gets 22\% of shifts, while attacking the next event causes a variation of 66\%. Finally, the percentage regarding the interval in the middle of the movie stimulus is 29\%. Next, \change{~\ref{JAMMovie/d}}{the right figure} highlights that the shift percentages of neutral event is 7\% when attacking 50\% of the neurons. In contrast, when the attack was executed at the beginning of the movie, the shifts percentage was 44\% and, finally, the shift percentage achieved when JAM was executed during the movie was 11\%. The difference between sets of neurons is appreciable, unlike the previous metric. However, once the high peak passes, the attack intervals start decreasing until remaining in the range of 50\%-60\% for both numbers of neurons.

\subsection{Impact study over the drifting gratings stimulus}

\begin{figure}[!htb]
\subfigure[\change{Total number of spikes when the attack is executed on 25\% of the neurons.}{Total number of spikes when the attack is executed on 25\% (left plot) and 50\% (right plot) of the neurons. Impacts under attack are emphasized using red numbers and gray shades.}]{\includegraphics[width=\textwidth]{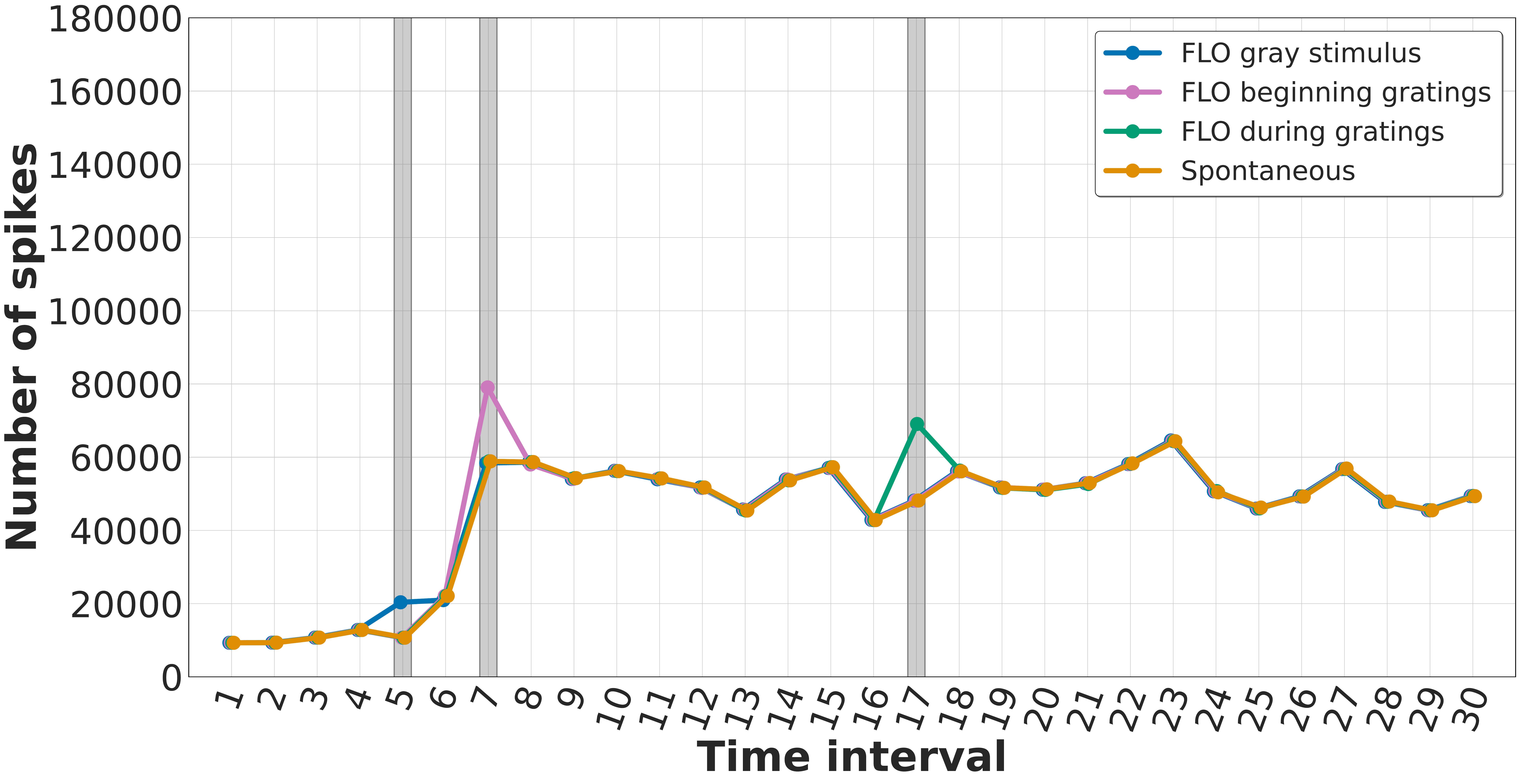}\label{FLOGratings/a}}
\subfigure[\change{Number of spikes of the impact over the 50\% of the neurons.}{Shift analysis of FLO when is executed on 25\% (left plot) and 50\% (right plot) of the neurons. Red numbers and gray shades marked the events affected by the attack.}]{\includegraphics[width=\textwidth]{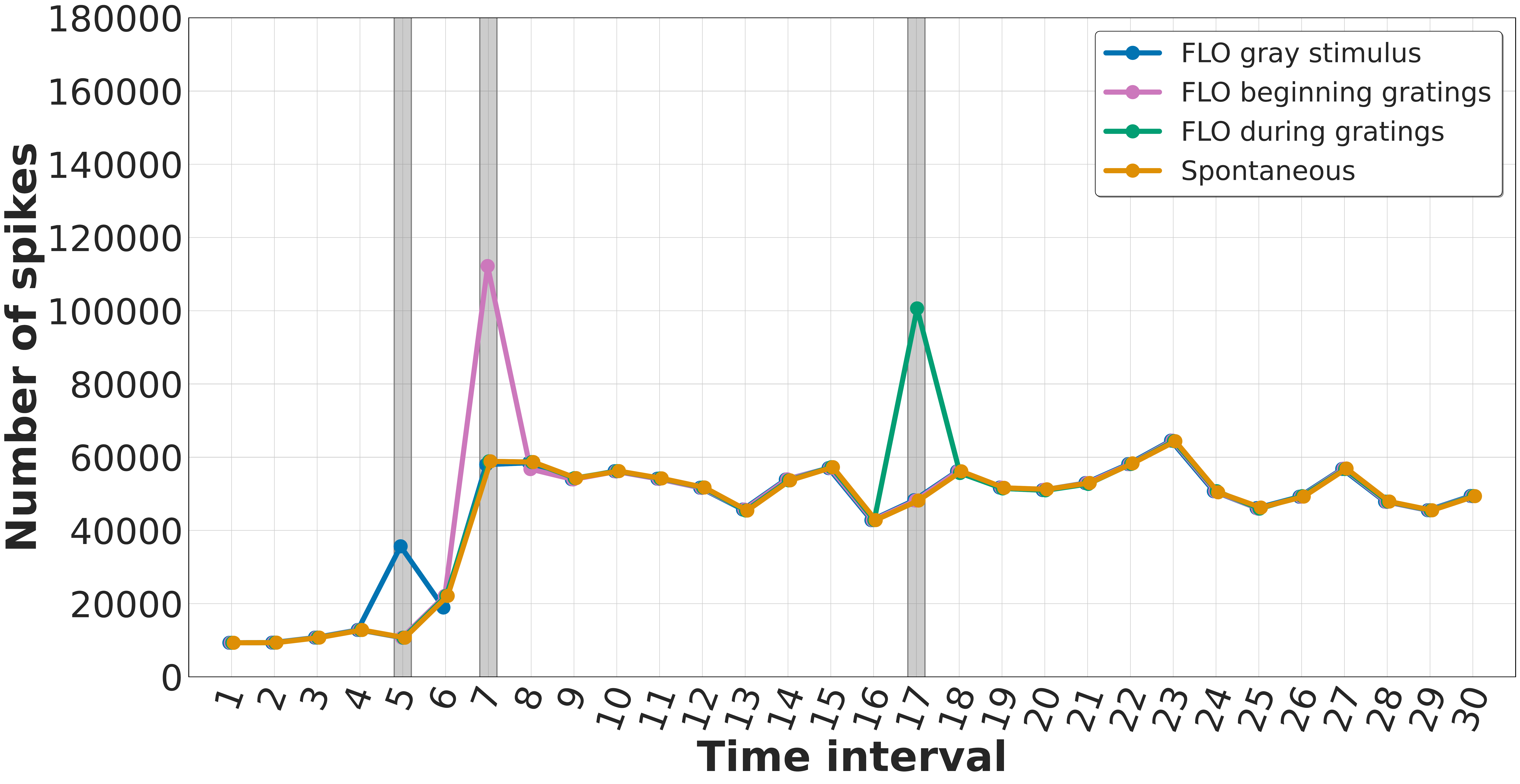}\label{FLOGratings/b}}
\caption{Analysis of the impact induced by FLO over the drifting gratings stimulus, focusing on metrics such as the number of spikes and shift percentage when targeting three different events. Besides, the experiments consider cases where 25\% and 50\% of neurons are affected. To asses variability, neurons were randomly selected ten times per experiment.}
\label{FLO Gratings}
\end{figure}

\figurename~\ref{FLO Gratings} shows the evolution of FLO executed over the drifting gratings stimulus. \addtxt{The left plot of }\figurename~\ref{FLOGratings/a} indicates that FLO causes an augmentation of 9,673 spikes (\change{90,43\%}{90.43\%}) when executed on the neutral event according to attacking 25\% of the neurons. The number of spikes when the attack is executed at the start of the drifting gratings event grows to 20,185 spikes (\change{34,29\%}{34.29\%}), even though attacking in the middle of the stimulus increases 20,899 more spikes (\change{43,39\%}{43.39\%}) than spontaneous activity. On the contrary, \change{~\ref{FLOGratings/b}}{the figure that represents the 50\% of the neurons in \figurename~\ref{FLOGratings/a}} indicates that attacking during the gray event increases the activity by 24,959 spikes (\change{233,35\%}{233.35\%}), whereas applying FLO at the beginning of the stimulus makes the number of spikes increases by 53,337 spikes (\change{90,61\%}{90.61\%}). Finally, in the middle of the stimulus, the number of spikes increments by 52,466 spikes (\change{108,92\%}{108.92\%}). Additionally, the recovery in the three attack instants takes around two or three intervals (200-300 ms). In both numbers of neurons under attack, the gray event is the most affected, as in the previous stimuli, due to the neurons generating few spikes in their normal behavior, causing a significant impact in the base activity when an attack is executed. However, the remaining events when the attack is executed on the 25\% of the neurons cause a lower number of spikes than the same events when FLO affected the 50\% of the neurons. Moreover, the results highlight that, during the stimulus, the neural reaction to the attack is greater compared to visualizing a neutral stimulus. 

Regarding the shift percentage metric, \addtxt{the plot showing 25\% of the neurons depicted in} \figurename~\ref{FLOGratings/b} indicates that the impact during the first instant is 69\% of shifts delayed. The value when the attack was executed at the beginning of the stimulus is 95\%, and, finally, the percentage obtained when FLO was used during the drifting gratings is 94\% of variation. In contrast, \change{~\ref{FLOGratings/d}}{the right figure} shows that the neutral event obtains 84\% of shifts; then, the instant corresponding to the beginning of the stimulus gets delayed by 98\%. Lastly, the impact obtained in the middle of the stimulus is 97\%. These results highlight high percentages in the whole simulation, remaining over 80\% except for the instant of the gray event when the attack affects 25\% of the neurons. Unlike the number of spikes metric where the neurons, after the attack, return to their normal activity, the shifts analysis shows that there is a great delay in each interval. Therefore, although the neurons generate a similar number of spikes as the spontaneous behavior, it does not mean that there is not a substantial delay in time. 

\begin{figure}[!htb]
\subfigure[\change{Number of spikes analysis of JAM when is executed on 25\% of the neurons.}{Number of spikes analysis of JAM when it is executed on 25\% (left plot) and 50\% (right plot) of the neurons. Red digits and gray tones remark on instants affected by attack.}]{\includegraphics[width=\textwidth]{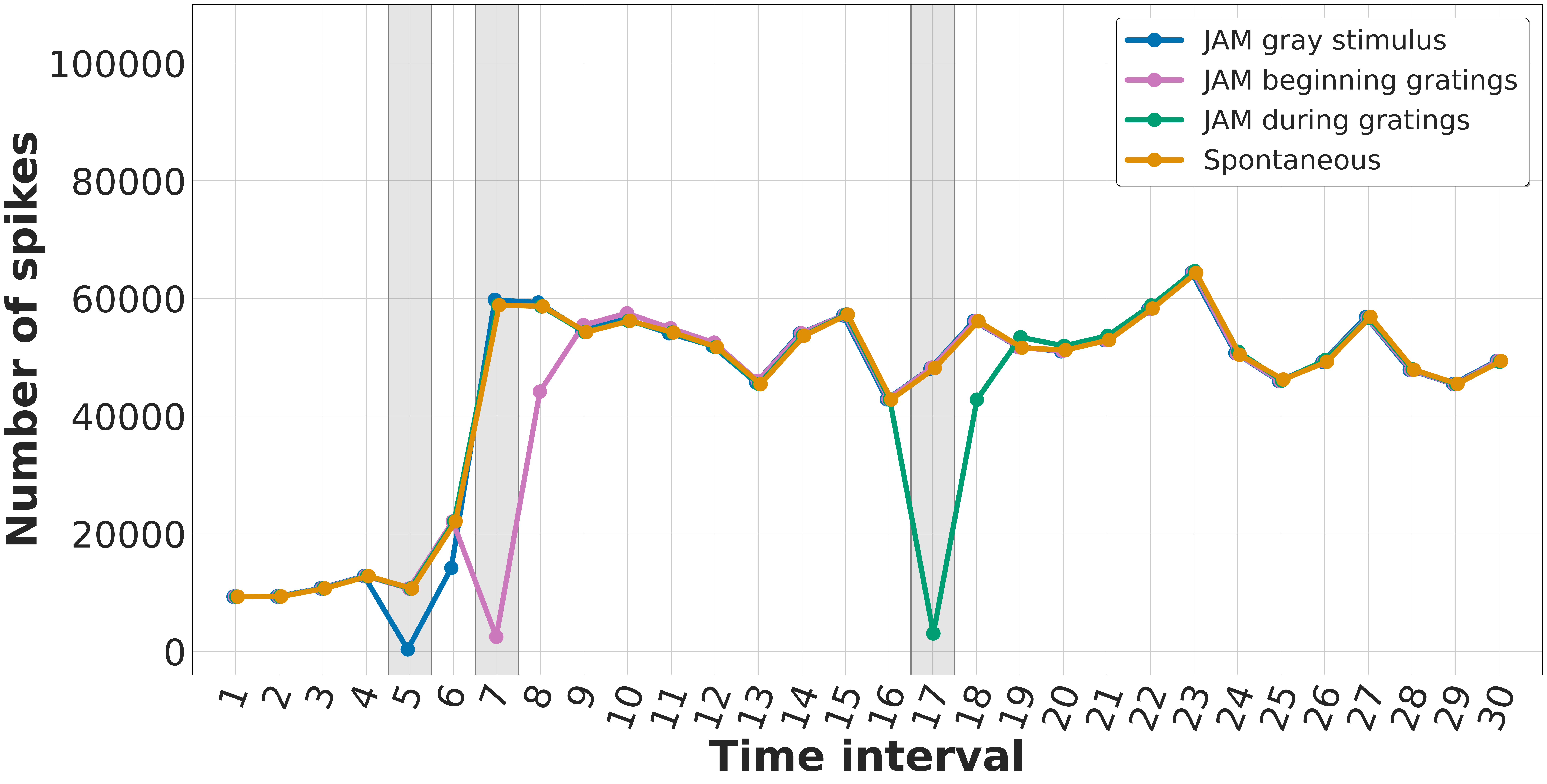}\label{JAMGratings/a}}
\subfigure[\change{Total number of spikes of the impact over the 50\% of the neurons.}{Percentages of shifts when the attack is executed on 25\% (left plot) and 50\% (right plot) of the neurons. Red indicators and gray-scale overlays highlighted the events under attack.}]{\includegraphics[width=\textwidth]{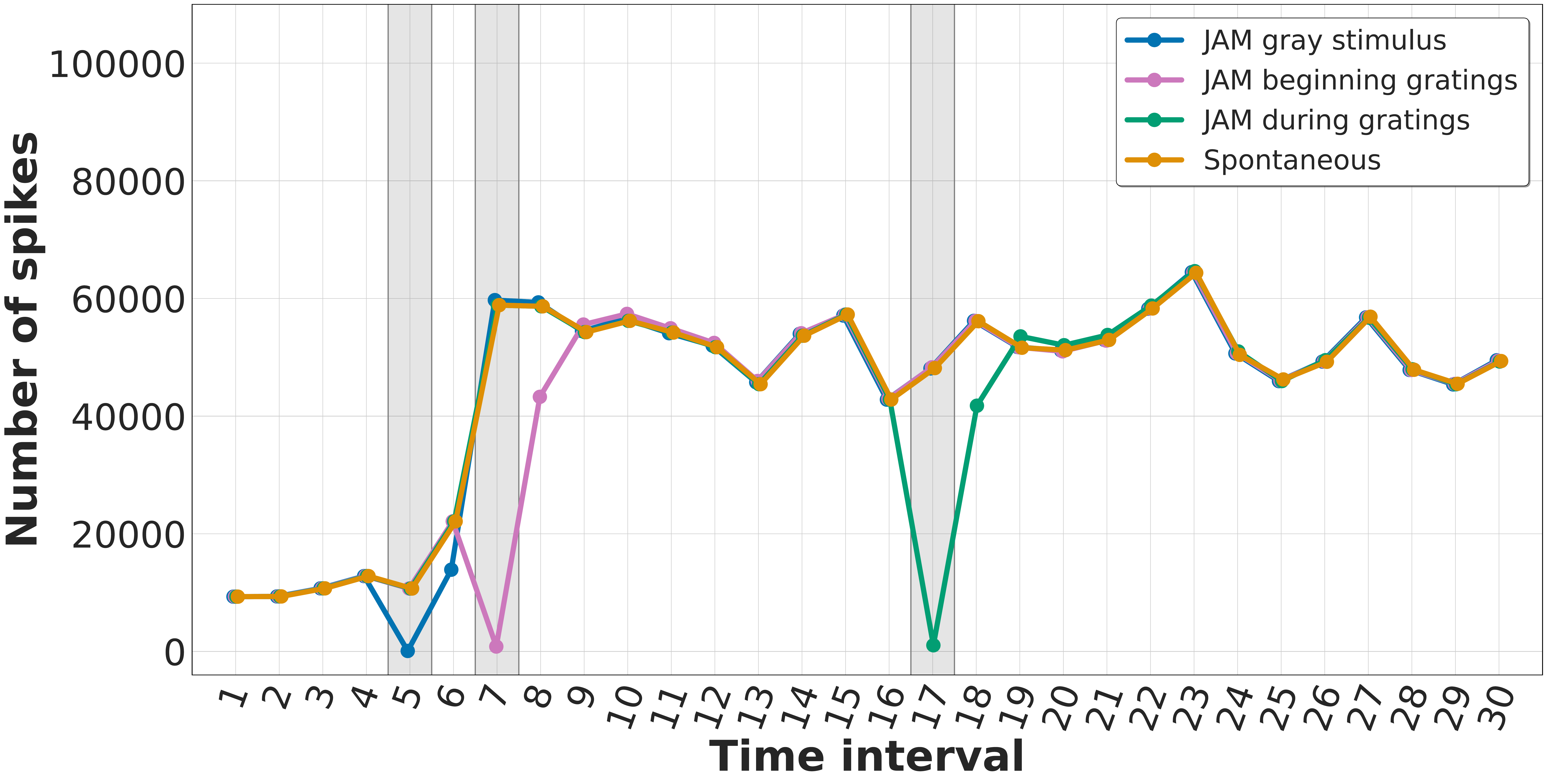}\label{JAMGratings/b}}
\caption{Evaluation of the impact induced by JAM over the drifting gratings stimulus considering the number of spikes and shift percentage metrics during attacks on three distinct events. The analysis differentiates cases with 25\% and 50\% of neurons affected, with neurons randomly selected ten times per experiment to study variability.}
\label{JAM Gratings}
\end{figure}

Finally, \figurename~\ref{JAM Gratings} presents the impact of JAM when it attacks the 25\% of the neurons and the 50\% of the neurons. On the one hand, \addtxt{the plot corresponding to 25\% of the neurons from} \figurename~\ref{JAMGratings/a} highlights that JAM decreases the number of spikes by 10,361 spikes (-96.87\%) in the gray event, while, in the second attack instant, the number of spikes has 56,387 spikes less (-95.79\%). In contrast, the diminishment obtained when JAM was used during the drifting gratings is 45,119 spikes (-93.67\%). On the other hand, \addtxt{the figure illustrating 50\% of the neurons presented in } \figurename~\ref{JAMGratings/b} depicts that, for the neutral event, the number of spikes decreases by 10,613 spikes (-99.22\%). Then, the results corresponding to the beginning of the event highlight 58,031 spikes less (-98.58\%), although the attack instant during the stimulus decrements by 47,121 spikes (-97.83\%). Besides, the neurons attacked in the instants corresponding to the gray event and in the middle of the stimulus return to their spontaneous activity after nine intervals (900 ms), while the neurons attacked at the beginning of drifting gratings require eight intervals (800 ms) to return to their normal behavior. In this case, the attack affects both sets of neurons equally, highlighting the low complexity to make a significant impact.

The percentage of shifts analysis is also divided by the number of neurons attacked. \addtxt{The left side of} \figurename~\ref{JAMGratings/b} indicates that the impact obtained in the neutral event is 44\%. Then, the percentage when the attack was executed at the beginning and during the drifting gratings stimulus is the same, 31\%. In contrast, \addtxt{the plot corresponding to 50\% of the neurons from }\figurename~\ref{JAMGratings/b} represents that attacking the gray event causes a delay of 17\%, whereas the following instants have the same percentage of shifts, 11\%. The results in the attack intervals, especially within the drifting gratings stimulus, suggest that the delay is the same regardless of when the attack is executed, where both attack instants have a similar evolution over the simulation. Nevertheless, in the remaining intervals, the percentage remains stable, over 80\%.

\section{Discussion}
\label{sec:discussion}
This section compares different relevant aspects of the results presented in Section \ref{sec:results}, such as the impact of neural cyberattacks on different visual stimuli, the difference between the cyberattacks according to the impact metrics, and the number of neurons attacked. Besides, this section compares these results with those existing in the literature. \addtxt{Furthermore, it presents an analysis either the effects on the brain, the mitigation strategies, and the scalability of neural cyberattacks.}

\subsection{Impact of neural cyberattacks on different visual stimuli}

All stimuli have the neutral event in common, which, as can be seen, is the most affected by both attacks, except when FLO is used in the movie stimulus. This situation could be explained as the result of attacking an event where the neurons generate few spikes in their normal behavior, causing a significant impact in the base activity. Moreover, the results of these gray events demonstrate that the effect of neural cyberattacks is consistent as these neutral events are included in different parts of the simulation and belong to different stimuli. Then, focusing on the remaining stimuli, the movie stimulus is more susceptible to attacks than the other stimuli, especially with FLO. After that, the results demonstrated that the drifting gratings stimulus is more vulnerable than ON-flash and OFF-flash, being the OFF-flash the most robust. Next, the propagation time is different according to the stimulus, especially in JAM. Thus, the events most affected by the propagation of JAM are the ON-flash and neutral events from the flash effect, while the movie stimulus and OFF-flash are less affected. Regarding the shifts analysis, the drifting gratings stimulus undergoes a longer delay than the others when both attacks are executed. On the contrary, the flash effect suffers less delay compared to the rest of the stimuli.

Besides, the impact observed over visual effects in motion, such as movie and drifting gratings stimuli, indicates that neural cyberattacks can cause a greater damage compared to fixed stimuli, such as the flash effect. Furthermore, the results highlight that attacking over bright events, such as the ON-flash, produces a larger impact compared to darker stimuli, like the OFF-flash.

\subsection{Impact comparison between cyberattacks}

It is necessary to analyze the impact of both cyberattacks individually to compare them. Focusing on FLO, and attending to the percentage of spikes increase compared to spontaneous behavior, the results indicate that the impact of FLO is highly dependent on the stimuli, getting a minimum percentage of increase of \change{12,74\%}{12.74\%} and a maximum of \change{94,64\%}{94.64\%} when 25\% of the neurons are attacked. In contrast, when FLO is used over 50\% of the neurons, the minimum percentage of increase is \change{32,83\%}{32.83\%}, and the maximum is \change{245,95\%}{245.95\%}. As can be observed, the minimum and maximum percentages of both sets of neurons coincide with the same events, being the OFF-flash the event with the lowest percentage and the middle of the movie stimulus the one with the highest percentage. In contrast, the overstimulation caused in terms of the number of spikes generated by FLO oscillates between 9,673 and 43,394 spikes for 25\% of the neurons, while in the 50\%, FLO oscillates between 24,959 and 96,798 spikes. These outcomes demonstrate that this attack fluctuated according to the event attacked. Secondly, the experimentation for JAM proves that this threat can cause a considerable damage regardless of the stimulus, causing a reduction of spikes between 93\% and 99\% compared to spontaneous signaling. Regarding the propagation of the effect caused by each attack, FLO propagates until 200-300 ms (2-3 intervals) depending on the stimuli evaluated, while JAM gets a propagation duration between 600 ms and 1,100 ms (6-11 intervals). In addition, the shifts analysis presents that both cyberattacks produced a substantial delay in the attack instants within the drifting gratings stimulus. Furthermore, the results depict that both attacks induce a delay of the neurons over the whole simulation, as can be seen in the shifts percentage figures, although the neurons seem to return to the spontaneous activity in the number of spikes metric.

Both cyberattacks have a significant impact, although JAM is more effective than FLO when randomly attacking the same amount of neurons. Besides, the shifts analysis indicates that JAM produces a longer delay compared to FLO on the whole simulation and for each stimulus. These results can be explained by the ability of JAM to suppress all activity from the neurons targeted, with independence of their inner behavior (excitatory or inhibitory). In contrast, performing a FLO attack over an inhibitory neuron could derive in the propagation of an inhibitory behavior over neurons establishing synaptic connections with the attacked neuron, thus alleviating the impact of FLO. 

\subsection{Comparative analysis of attacked number of neurons}

The impact comparison in terms of the number of neurons attacked, based on the results presented in Section \ref{sec:results}, indicates that FLO has a greater effect when executed over 50\% of the neurons regarding both the number of spikes and percentage of shifts metrics. On the contrary, JAM causes a similar impact in both sets of neurons attending to the number of spikes metric; therefore, the results indicate that it is not necessary to attack a high number of neurons to produce a notorious impact. Focusing on the propagation of the attack considering the number of spikes, there is not a distinguished difference between both sets of neurons with independence of the attack. Regarding the percentage of shifts when JAM is executed, the only difference between these sets of neurons is during the attack interval, being the remaining intervals similar for both numbers of neurons.  

\subsection{Comparison between the literature and the current research}

After discussing the results from this research, this manuscript compares the experiments proposed with those done in the literature. In the topology of 450 neurons corresponding to layer four of V1, presented by López Madejska et al. \cite{Lopez_Madejska:realitic_topology:2024}, the authors employed as input static values represented as LGN spikes, whereas this work used a more complex topology, which represents all six layers from V1, comprised of more than 230,000 neurons with visual realistic stimuli as inputs: flash effect, a movie composed by three scenes and a drifting gratings stimulus of 90 degrees. Regarding the neuronal model used, the topology from this work uses morpho-electrical characteristics with greater details than those from the literature. 

Regarding FLO, the literature conducted two experiments: 1) study of the impact of using different voltages to attack, and 2) the effect of attacking different numbers of neurons. Focusing on the first, the authors concluded that the higher the voltage, the more effective the attack. Therefore, this research has utilized parameters already documented by López Madejska et al. \cite{Lopez_Madejska:realitic_topology:2024} and López Bernal et al., \cite{Lopez_Bernal:cyberattacks_implants:2020}. Concerning the second experiment, attacking a higher number of neurons derived in a larger impact, obtaining the same results in the present work. Moving to JAM, the literature concludes that the higher the number of neurons attacked, the higher the impact, observing an incremental impact when augmenting the neurons attacked, testing several configurations between 50 and 450 neurons. However, the present work concludes that attacking 25\% and 50\% of the neurons generate a similar impact, highlighting that it is not necessary to attack a great number of neurons to produce a considerable impact over normal behavior. This difference can be explained by the complexity of the topology used and the larger degree of detail of the models used, although further research is needed. Regarding the propagation time, in the literature, FLO lasts around 500 ms, whereas in this research is between 200 and 300 ms in the three stimuli. On the contrary, JAM has a propagation duration of 600 ms in the literature; however, in this work, the effect of JAM persists between 600 and 1,100 ms according to the event for the flash effect. Then, 700 ms or 600 ms for the movie stimulus, and around 800-900 ms for the drifting gratings stimulus. It is important to highlight that these durations depend on the attack instant. 

Despite the advances conducted in this work, there are several limitations worth considering. First, the impact analysis performed does not differentiate between attacking excitatory and inhibitory neurons, the neuronal populations under attack, or the layers targeted. Thus, further analysis  could improve the knowledge of the impact of neural cyberattacks on neuronal behavior. Moreover, the study of the cyberattacks performed only considers the activity of the neurons, without considering other dimensions of neural networks, such as altering connection between neurons. Furthermore, this research evaluated visual stimuli in black and white without the evaluation of the possible behavior of neurons in response to attacking during a stimulus with color. Besides, the model used covers a small portion of the primary visual cortex (\SI{845}{\micro\metre} radius) of the mice, reflecting the need to use more complex topologies closer to complexity of the human brain. However, there at not, at the moment, more detailed neuronal reconstructions modeling the visual cortex. 

\addtxt{\subsection{Analysis of the effects on the brain and mitigation strategies.}}
\addtxt{The execution path for neural cyberattacks in the real world begins with the exploitation of vulnerabilities existing in a neurostimulation device. In this regard, López Bernal et al. \cite{Lopez_Bernal:cyberattacks_implants:2020} identified several cybersecurity flaws in emerging neurostimulation implants, including those used in Neuralink. These vulnerabilities enable an attacker to gain control over the device and manipulate its operation. The consequences of unauthorized access include alterations in normal brain activity, which can result in adverse effects on the patient. In this context, the literature has documented that inadequate or malicious neurostimulation can lead to severe physical and psychological consequences in the brain. This risk is particularly critical in devices such as BCIs, where the artificial generation or alteration of brain signals may result in serious health risks, as described by López Bernal et al. \cite{Lopez_Bernal:cyberBCI:2021}. For instance, during neurostimulation therapies such as those used for Parkinson's disease, cyberattacks could induce the inability to initiate voluntary movement, memory difficulties, or general weakness as neurological side effects. Psychological consequences may include anxiety, confusion, or depression. In extreme cases, these attacks could even provoke death. In the context of visual neurostimulation, potential effects might include temporary or permanent blindness, inability to perceive motion, or inability to recognize objects}

\addtxt{Focusing on possible protection countermeasures to detect neural cyberattacks and mitigate their effects, the following mechanisms could be applied. Firstly, as a detection technique, artificial intelligence can be used to identify anomalies by identifying unusual patterns or significant deviations in neural activity. These AI-based systems enhance the capability to detect subtle changes that may indicate the presence of an attack. Additionally, implementing real-time alert systems, whether directed at human operators or automated control systems, aims to facilitate rapid intervention in the event of detected threats. Furthermore, logging mechanisms can be integrated to systematically record spike activity, input stimuli, external perturbations, and timing information for the detection of attacks. Secondly, within mitigation measures, the logging technique, as mentioned previously, provides a foundation for post-attack analysis, enabling traceability, correlation between attack events and their effects, and support for the development of future mitigation strategies based on the actions taken in response to the attacks. Besides, generating reports on the effects and scope of neural cyberattacks would contribute to minimizing their impact and improving the security of BCIs. To interrupt the propagation of the attack, temporarily stopping BCI operation can prevent the escalation of the attack impact before it compromises the patient's health. Finally, conducting a session to educate both clinicians and patients about the formative sessions could help reduce the risks of neurostimulation BCIs and their consequences, as humans are usually the weakest element in security systems. }

\addtxt{\subsection{Evaluating the Scalability of Neural Cyberattacks}}
\addtxt{In both real and simulation scenarios, there is direct access to individual neurons, either through multiple electrodes in the case of BCIs or through the neuronal topology used in the simulation. Therefore, the methodology used to execute the attacks, applied in this case to V1, can be extended to other brain regions, such as the motor or prefrontal cortex, since the approach is generalizable. However, changing the target of the attack introduces several new challenges. In real BCIs, the intruder must understand the functional characteristics of the neurons in the new cortical region, including their connectivity patterns, firing dynamics, and general behavior. Each brain area has distinct structural and functional properties, which complicates the design and deployment of effective attacks. In simulation scenarios, the attack code must be adapted to the new realistic neural topology once the corresponding region has been analyzed and modeled. This adjustment includes reconfiguring the input of the model, adjusting attack parameters such as the number of targeted neurons or the attack interval, and validating the results. In both real and simulated contexts, it becomes necessary to evaluate how the new area responds to cyberattacks, analyze the specific vulnerabilities it may present, and perform comparative studies between results.}
\section{Conclusion}
\label{sec:conclusions}
This work analyzes the impact of two neural cyberattacks, FLO and JAM, implemented on a complex reconstruction of the primary visual cortex from mice, comprised of around 230,000 neurons. This topology is also formed by six layers of V1 with different neuronal populations and behaviors. Besides, this network utilized three realistic visual stimuli: a flash effect, a movie, and drifting gratings. This manuscript studied different attack instants over each stimulus, the optimal background and LGN trials, and parameters such as the number of neurons, the voltages, and the number of times the experiments are executed to conduct these attacks. Once the experiments are completed, two impact metrics are used: the number of spikes and the percentage of shifts. The results indicated that the movie stimulus is more vulnerable to cyberattacks than the remaining stimuli, while the black event from the flash effect is the most robust. Therefore, performing the attacks on bright and complex (e.g., in motion) stimuli produces a greater impact. Additionally, attacking during an event where the neurons generate few spikes in their normal behavior produces a significant increase in the base activity compared to situations that naturally present larger neural activity.

Although FLO and JAM have a significant impact, FLO is determined by the stimulus and the attack instant, whereas JAM disrupts neuronal activity independently of the event and the attack interval. Thus, JAM is more damaging and produces a longer delay in most experiments, regarding the number of spikes. Additionally, the percentage of shifts metric indicates that both attacks produce a delay of spikes in time compared to spontaneous signaling in both attacks. In the same way, the propagation of the effect caused by the attacks changes according to the attack instant. FLO has a propagation duration between 200 and 300 ms for all the stimuli, whereas JAM has a different duration according to the stimuli attacked, with 600 ms the shorter and 1100 ms the longest. Therefore, the effect caused by JAM is more persistent than FLO. Moreover, JAM does not require to attack a considerable number of neurons to cause considerable damage. On the contrary, in FLO, the higher the number of neurons attacked, the higher the impact. 

Future work could explore the identification of new cyberattacks focused on other model components, such as the connections between neurons. \addtxt{Besides, implementing the remaining six neural cyberattacks from the literature would enable a more comprehensive analysis of the effects these types of attacks can have on neural activity.} In the same way, there is a need for works testing the impact and effectiveness of neural cyberattacks on excitatory and inhibitory neurons, over different layers of V1, and distinct neuronal population. In addition, further work is necessary to evaluate the reaction of neurons in the face of color images and static gratings to compare with the movie and drifting grating stimulus used in this paper, respectively. New research also needs to assess other impact metrics to analyze other aspects of neural behavior\addtxt{, such as synchronization and group dynamics, including metrics like mean firing rate and population synchrony.} \addtxt{Moreover, metrics related to visual discrimination performance, such as reaction time, which measures the time required for neurons to respond to the presentation of a stimulus, would be interesting to use in our study under the effects of the attacks.} Finally, to obtain novel conclusions regarding the impact that these cyberattacks could generate in the real world, it is necessary to keep a close collaboration with neuroscience specialists to evaluate and recreate certain attack situations in real life, such as temporary blindness, or generate stimuli where there were none. 

\section*{Declaration of Competing Interest}
The authors declare that they have no known competing financial interests or personal relationships that could have appeared to influence the work reported in this paper. 

\section*{CRediT authorship contribution statement}

\textbf{Victoria Magdalena L\'opez Madejska.} Methodology, Conceptualization, Investigation, Writing - original draft, Data curation, Software, Visualization.
\textbf{Sergio L\'opez Bernal.} Methodology, Conceptualization, Writing - Review \& Editing. 
\textbf{Gregorio Mart\'inez P\'erez.} Supervision, Project administration, Funding acquisition.
\textbf{Alberto Huertas Celdr\'an.} Methodology, Conceptualization, Funding acquisition, Writing - Review \& Editing.

\section*{Acknowledgment}
This work has been partially funded by (a) the strategic project "Development of Professionals and Researchers in Cybersecurity, Cyberdefense and Data Science (CDL-TALENTUM)" from the Spanish National Institute of Cybersecurity (INCIBE) and by the Recovery, Transformation and Resilience Plan, Next Generation EU, (b) the Swiss Federal Office for Defense Procurement (armasuisse) with the CyberTracer (CYD-C-2020003) project, and (c) the University of Z\"urich UZH.

%
%
\bibliographystyle{IEEEtran}
\bibliography{biblio}

\end{document}